\documentclass[onecolumn, draftcls, 12pt]{IEEEtran}
\usepackage{amsfonts}
\usepackage{mathrsfs}
\usepackage{amssymb,amsmath}

\usepackage{algorithm}
\usepackage{algorithmic,mathrsfs}
\usepackage{amsmath,amssymb,epsfig,graphics,subfigure}
\usepackage{theorem}
\usepackage{array,color}
\usepackage[compress]{cite}

\theoremheaderfont{\normalfont\bfseries}

\begin{document}

\title{A General Robust Linear Transceiver Design for Multi-Hop Amplify-and-Forward MIMO Relaying Systems}

\author{Chengwen Xing, Zesong Fei, Shaodan Ma, Yik-Chung Wu and H. Vincent Poor
 \thanks{Chengwen Xing and Zesong Fei are with the School of Information and Electronics, Beijing Institute of Technology, Beijing, China. Phone : (86)1068911841, Fax : (86)1068912615, Email : xingchengwen@gmail.com, feizesong@bit.edu.cn.}
\thanks{Shaodan Ma is with the Department of Electrical and Computer Engineering, University of Macau, Macao. Email: shaodanma@umac.mo.}
 \thanks{Yik-Chung Wu is with the Department of Electrical and Electronic Engineering, The University of Hong Kong, Hong Kong. Email : ycwu@eee.hku.hk. }
 \thanks{H. Vincent Poor is with the Department of Electrical Engineering, Princeton University, Princeton, NJ 08544 USA. Email: poor@princeton.edu.}
 \thanks{The material in this paper was partially presented at the International Conference on Wireless Communications and Signal Processing (WCSP), Nanjing, China, Nov. 2011. $^*$The corresponding author is Shaodan Ma.}
}

\maketitle

\begin{abstract}

In this paper, linear transceiver design for multi-hop
amplify-and-forward (AF) multiple-input multiple-out (MIMO) relaying systems with Gaussian distributed channel estimation errors is investigated. Commonly used transceiver design criteria including weighted mean-square-error (MSE) minimization, capacity maximization, worst-MSE$/$MAX-MSE minimization and weighted sum-rate maximization, are considered and unified into a single matrix-variate optimization problem. A general robust design algorithm is proposed to solve the unified problem. Specifically, by exploiting majorization theory and properties of matrix-variate functions, the optimal structure of the robust transceiver is derived when either the covariance matrix of channel estimation errors seen from the transmitter side or the corresponding covariance
matrix seen from the receiver side is proportional to an identity matrix. Based on the optimal structure, the original transceiver design problems are reduced to much simpler problems with only scalar variables whose solutions are readily obtained by iterative water-filling algorithm. A number of existing transceiver design algorithms are found to be special cases of the proposed solution. The differences between our work and the existing related work are also discussed in detail. The performance advantages of the proposed robust designs are demonstrated by simulation results.


\end{abstract}

\section{Introduction}
\label{sect:intro}

 With significant potential to enable the emerging requirements for high speed ubiquitous wireless communications, cooperative communications has been adopted as one of the key components in future wireless communication standards such as long term evolution (LTE), international mobile telecommunications-advanced (IMT-Advanced), the Winner project, etc. Specifically, these developments involve the deployment of relays to enhance the coverage of base stations and to improve the communication quality of wireless links \cite{Laneman04}. In general, relays can adopt different relaying strategies, e.g., amplify-and-forward (AF), decode-and-forward (DF) and compress-and-forward (CF). Among these relaying strategies, the AF scheme is the most attractive for practical implementation due to its low complexity and independence of the underlying modulation. On the other hand, it is well-established that employing multiple antennas provides spatial diversity and multiplexing gain in a wireless communication system. It is straightforward to combine AF transmission with multi-input multi-output (MIMO) systems so that the virtues of both techniques can be obtained. The resulting system (termed an AF MIMO relaying system) has attracted considerable interest \cite{Jin2010} in recent years.

 Transceiver design for AF MIMO relaying systems, which refers to the design of source precoder, relay amplifier and receiver equalizer, has been widely discussed in the literature \cite{Medina07,Tang07,Guan08,Mo09,Rong09,Tseng09,Li09,
 Rong2009TWC,Xing10,Xing1012,Rong20115,XingGlobecom,XingICSPCC,Chalise10}. Generally speaking, transceiver design varies from system to system and depends heavily on the design criteria and objectives. The most commonly used criteria are capacity maximization \cite{Tang07,Rong09,Medina07} and data mean-square-error (MSE) minimization \cite{Guan08,Mo09,Tseng09,Rong09}. Usually these two criteria are contradictory to each other and call for different algorithms to solve the optimization problems. Interestingly, in \cite{Rong09} a unified framework which is applicable to both capacity maximization and MSE minimization is proposed for transceiver design in dual-hop AF MIMO relay systems. Since multi-hop AF transmission is a promising technique to increase the coverage of a transmitter, transceiver design for a multi-hop system is further investigated in \cite{Rong2009TWC}. It reveals that optimal solutions for both capacity and MSE criteria in a multi-hop system should have diagonal structures. However, in most of the previous works on transceiver design including \cite{Rong09} and \cite{Rong2009TWC}, channel state information (CSI) is assumed to be perfectly known/estimated. This is difficult to achieve in practice and channel estimation errors are inevitable due to limited training and quantization operation, resulting in significant performance degradation. In order to mitigate the performance degradation, such channel estimation errors are necessary to be taken into account in the transceiver design process. This kind of transceiver is called robust transceiver. It has been shown in \cite{Ding09} and \cite{Ding10} that robust transceiver design is essentially different from the transceiver design with perfect CSI. It is more challenging and different algorithms are required to solve the challenging robust design problem.

 In general, channel estimation errors can be modeled in two different ways: norm-bounded errors with known error bound and random errors with certain distribution. Correspondingly, robust transceiver designs can also be classified into two main categories: worst-case robust design for norm-bounded errors \cite{Wang2010} and Bayesian robust design for randomly distributed errors \cite{Zhang08}. For linear channel estimators, the estimation errors can be accurately modeled as being random with a Gaussian distribution \cite{Xing1012}. Under this kind of Gaussian estimation errors, Bayesian robust transceiver design for dual-hop AF relaying systems has been investigated in \cite{Rong20115} and solutions for capacity maximization and MSE minimization respectively are proposed by implicitly approximating a design-variable dependent covariance matrix (the matrix ${\bf{A}}$ in \cite{Rong20115}) as being constant. Since the approximation is tight only when the covariance matrix of channel estimation errors seen from the receiver side is proportional to an identity matrix, the proposed solutions are sub-optimal for general cases. In \cite{Xing1012} and \cite{Xing10}, Bayesian robust transceiver design targeting at weighted MSE minimization is discussed for dual-hop AF relaying systems and an optimal solution is found without considering the source precoder. The optimality of the proposed solution is proved to hold under a wide range of cases, i.e., when either the covariance matrix of channel estimation errors from the transmitter side or the corresponding covariance matrix seen from the receiver side is proportional to an identity matrix. These works have been extended to systems with source precoder design and an iterative algorithm has been proposed to find a good solution without guaranteed optimality \cite{XingGlobecom}. Similarly, the robust transceiver design for maximizing mutual information rate for dual-hop AF relaying systems under Gaussian channel estimation errors at all nodes has been investigated in \cite{XingICSPCC} and a solution without global optimality is proposed with an iterative algorithm. Unfortunately, the aforementioned algorithms are applicable only to dual-hop AF systems and their extension to multi-hop AF systems is by no means straightforward as shown in \cite{Rong2009TWC}.


In this paper, we investigate robust transceiver design for a general \textbf{\textsl{multi-hop}} AF relaying system with Gaussian distributed channel uncertainties. The robust design problem is significantly different from that in the literature and is challenging due to the existence of random channel uncertainties and the complexity of the multihop system. A number of widely used design criteria including weighted MSE minimization, capacity maximization, worst case MSE minimization, and weighted sum-rate maximization are considered and their corresponding robust design problems are unified into one matrix-variate optimization problem. A general robust design algorithm is proposed to solve the unified problem, i.e., to jointly design the precoder at the source, multiple forwarding matrices at the relays, and the equalizer at the destination. Specifically, the structure of the optimal solution for the unified problem is derived based on majorization theory \cite{Palomar2006,Jorswieck07} and properties of matrix-monotone functions \cite{Jorswieck07}. It is demonstrated that the derived optimal structure is significantly different from its counterpart with perfect CSI \cite{Rong2009TWC} and its optimality holds under a wide range of cases, i.e., when either the covariance matrix of channel estimation errors seen from the transmitter side or the corresponding covariance matrix seen from the receiver side is proportional to an identity matrix. With the optimal structure, the robust design problem is simplified into a design problem with only scalar variables. An iterative water-filling algorithm is then proposed to obtain the remaining unknown parameters in the transceiver. The performance of the proposed robust designs is finally corroborated by simulation results. In addition, it is shown that the proposed solutions cover some existing transceiver design solutions as special cases.

The rest of the paper is organized as follows. In Section~\ref{set: system model}, the signal model for a multi-hop AF system is introduced. Then a unified robust transceiver design problem applicable to weighted MSE minimization, capacity maximization, MAX-MSE minimization and weighted sum-rate maximization, is formulated in Section~\ref{sect:problem formulation}. In Section~\ref{sect:structure}, the optimal structure for the robust transceiver is derived and the unified transceiver design problem is reduced to a problem of finding a set of diagonal matrices, which can be solved by an iterative water-filling algorithm. The performance of the proposed robust designs is demonstrated by simulation results in Section~\ref{sect:simulation}. Finally, this paper is concluded in Section~\ref{sect:conclusion}.

The following notation is used throughout this paper. Boldface
lowercase letters denote vectors, while boldface uppercase letters
denote matrices. The notation ${\bf{Z}}^{\rm{H}}$ denotes the
Hermitian of the matrix ${\bf{Z}}$, and ${\rm{Tr}}({\bf{Z}})$ is the
trace of the matrix ${\bf{Z}}$. The symbol ${\bf{I}}_{M}$ denotes the
$M \times M$ identity matrix, while ${\bf{0}}_{M,N}$ denotes the $M
\times N$ all zero matrix. The notation ${\bf{Z}}^{1/2}$ is the
Hermitian square root of the positive semidefinite matrix
${\bf{Z}}$, such that ${\bf{Z}}^{1/2}{\bf{Z}}^{1/2}={\bf{Z}}$ and
${\bf{Z}}^{1/2}$ is also a Hermitian matrix. The symbol $\lambda_i({\bf{Z}})$ represents the $i^{\rm{th}}$ largest eigenvalue of ${\bf{Z}}$. The symbol $\otimes$ denotes
the Kronecker product. For two Hermitian matrices, ${\bf{C}} \succeq
{\bf{D}}$ means that ${\bf{C}}-{\bf{D}}$ is a positive semi-definite
matrix. The symbol ${\boldsymbol \Lambda} \searrow $ represents a rectangular diagonal
matrix with decreasing diagonal elements.

\section{System Model}
\label{set: system model}

\begin{figure}[!ht]
\centering
\includegraphics[width=.8\textwidth]{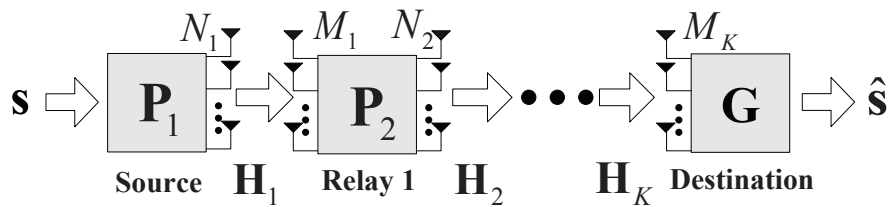}
\caption{A multi-hop amplify-and-forward MIMO relaying system.}\label{fig:1}
\end{figure}

In this paper, a multi-hop AF MIMO relaying system is considered. As shown in Fig.~\ref{fig:1}, one source with $N_1$ antennas wants to communicate with a destination with $M_{K}$ antennas through $K-1$ relays. The $k^{\rm{th}}$ relay has $M_{k}$ receive
antennas and $N_{k+1}$ transmit antennas. It is obvious that the dual-hop AF MIMO relaying system is a special case of this configuration when $K=2$.

At the source, an $N\times 1$ data vector s with  covariance matrix ${\bf{R}}_{\bf{s}} = {\mathbb{E}}\{{\bf{s}}{\bf{s}}^{\rm H}\} = {\bf{I}}_N$ is transmitted after being precoded by a precoder matrix ${\bf{P}}_1$. The received signal ${\bf{x}}_1$ at the first relay is ${\bf{x}}_1= {\bf{H}}_{1}{\bf{P}}_1{\bf{s}}+{\bf{n}}_1$ where ${\bf{H}}_{1}$ is the MIMO channel matrix between the source and the first relay, and ${\bf{n}}_1$ is an additive Gaussian noise vector at the first relay with zero mean and
covariance matrix ${\bf{R}}_{n_1}=\sigma_{n_1}^2{\bf{I}}_{M_1}$.

At the first relay, the received signal ${\bf{x}}_1$ is multiplied by a
forwarding matrix ${\bf{P}}_2$ and then the resulting signal is
transmitted to the second relay. The received signal ${\bf{x}}_2$ at
the second relay is ${\bf{x}}_2={\bf{H}}_{2}{\bf{P}}_2{\bf{x}}_1+{\bf{n}}_2$, where ${\bf{H}}_{2}$ is the MIMO channel matrix between the first relay
and the second relay, and ${\bf{n}}_2$ is an additive Gaussian noise
vector at the second relay with zero mean and covariance matrix ${\bf{R}}_{n_2}=\sigma_{n_2}^2{\bf{I}}_{M_2}$. Similarly, the received signal at the $k^{\rm{th}}$ relay can be written as
\begin{align}
{\bf{x}}_{k}={\bf{H}}_{k}{\bf{P}}_{k}{\bf{x}}_{k-1}+{\bf{n}}_{k}
\end{align}where ${\bf{H}}_k$ is the channel matrix for the $k^{\rm{th}}$ hop, and ${\bf{n}}_{k}$ is an additive Gaussian noise vector with zero mean and covariance matrix ${\bf{R}}_{n_{k}}=\sigma_{n_k}^2{\bf{I}}_{M_{k}}$.

Finally, for a $K$-hop AF MIMO relaying system, the received signal at the destination is
\begin{align}
{\bf{y}} = [{\prod_{k=1}^K}{\bf{H}}_{k}{\bf{P}}_k]{\bf{s}}  + \sum_{k=1}^{K-1}\{ [\prod_{l={k+1}}^K{\bf{H}}_{l}{\bf{P}}_l]{\bf{n}}_k
\}+{\bf{n}}_K,
\end{align}where ${\prod_{k=1}^K}{\bf{Z}}_k$ denotes ${\bf{Z}}_K\times \cdots \times {\bf{Z}}_1$. It is generally assumed that
$N_k$ and $M_k$ are greater than or equal to $N$ in order to guarantee that the transmitted data ${\bf{s}}$ can be recovered at the destination \cite{Guan08}.

In practical systems, because of limited length of training sequences, channel estimation errors are inevitable.
With channel estimation errors, the channel matrix can be written as
\begin{align}
 {\bf{H}}_{k}&={\bf{\bar H}}_{k}+\Delta{\bf{H}}_{k},
\end{align}where ${\bf{\bar H}}_{k}$ is the estimated channel matrix in the $k^{\rm{th}}$ hop and $\Delta{\bf{H}}_{k}$ is the corresponding channel estimation error matrix whose elements are zero mean Gaussian random variables. Moreover, the $M_k \times
N_k$ matrix $\Delta{\bf{H}}_{k}$ can be decomposed using the widely used Kronecker model as
$\Delta{\bf{H}}_{k}=
{\boldsymbol{\Sigma}}_{k}^{{1}/{2}}{\bf{H}}_{W,k}
{\boldsymbol{\Psi}}_{k}^{{1}/{2}}$ \cite{Ding09,Ding10,Zhang08,Xing10,Xing1012,Chalise10}. The elements of the $M_k \times N_k$ matrix
${\bf{H}}_{W,k}$ are independent and identically distributed
(i.i.d.) Gaussian random variables with zero means and unit variances. The specific properties of the row correlation matrix ${\boldsymbol \Sigma}_k$ and the  column correlation matrix ${\boldsymbol \Psi}_k$ are determined by the training sequences and channel estimators being used \cite{Ding09,Xing1012}. Note that ${\boldsymbol \Sigma}_k$ and ${\boldsymbol \Psi}_k$ correspond to the covariance matrices of the channel estimation errors seen from the transmitter and receiver sides, respectively.

At the destination, a linear equalizer ${\bf{G}}$ is employed to
detect the desired data vector ${\bf{s}}$. The resulting data MSE
matrix equals to $
{\boldsymbol{\Phi}}({\bf{G}},\{{\bf{P}}_k\}_{k=1}^K)=\mathbb{E}\{({\bf{G}}{\bf{y}}-{\bf{s}})({\bf{G}}{\bf{y}}-{\bf{s}})^{\rm{H}}\} $, where the expectation is taken with respect to random data, channel estimation errors, and noise\footnote{Here the channel estimation errors are assumed unknown at all the nodes. The data MSE matrix at the receiver should thus be computed by taking expectation against all the unknown random variables including data, noise and channel estimation errors.}.
Following a similar derivation in dual-hop systems \cite{Xing10}, the MSE matrix is derived to be
\begin{align}
\label{MSE_final}
{\boldsymbol {\Phi}}({\bf{G}},\{{\bf{P}}_k\}_{k=1}^K)=& \mathbb{E}\{({\bf{G}}{\bf{y}}-{\bf{s}})
({\bf{G}}{\bf{y}}-{\bf{s}})^{\rm{H}}\}\nonumber \\ =& {\bf{G}}[{\bf{\bar
H}}_{K}{\bf{P}}_K{\bf{R}}_{{\bf{x}}_{K-1}}{\bf{P}}_K^{\rm{H}} {\bf{\bar
H}}_{K}^{\rm{H}}+ {\rm{Tr}}({\bf{P}}_K{\bf{R}}_{{\bf{x}}_{K-1}} {\bf{P}}_K^{\rm{H}}
{\boldsymbol\Psi} _{K}
   ){\boldsymbol \Sigma}_{K} \nonumber \\
&+{\bf{R}}_{n_K}]{\bf{G}}^{\rm{H}}+ {\bf{I}}_N-[\prod_{k=1}^K{\bf{\bar H}}_{k}{\bf{P}}_k]^{\rm{H}}{\bf{G}}^{\rm{H}}-
{\bf{G}}[\prod_{k=1}^K{\bf{\bar H}}_{k}{\bf{P}}_k],
\end{align}where the received signal covariance matrix ${\bf{R}}_{{\bf{x}}_k}$ at the $k^{\rm{th}}$ relay satisfies the following recursive formula:
\begin{align}
\label{R_x}
{\bf{R}}_{{\bf{x}}_k}&= {\bf{\bar H}}_{k}{\bf{P}}_k{\bf{R}}_{{\bf{x}}_{k-1}}{\bf{P}}_k^{\rm{H}}{\bf{\bar H}}_{k}^{\rm{H}}+ { {\rm{Tr}}({\bf{P}}_k{\bf{R}}_{{\bf{x}}_{k-1}} {\bf{P}}_k^{\rm{H}}
{\boldsymbol\Psi} _{k}
   ){\boldsymbol\Sigma} _{k}+{\bf{R}}_{n_k}},
\end{align}and ${\bf{R}}_{\bf{x}_0}={\bf{R}}_{\bf{s}}={\bf{I}}_N$ represents the signal covariance matrix at the source.

\section{Transceiver Design Problems}
\label{sect:problem formulation}

\subsection{Objective Functions}
\label{sect:objectives}

There are various performance metrics for transceiver design. In the following, four widely used metrics are discussed.

(1) \textit{Weighted MSE}: With the data MSE defined in (\ref{MSE_final}), weighted MSE can be directly written as
\begin{align}
\label{Obj 1}
\text{Obj 1:} \ \ {\rm{Tr}}[{\bf{W}}{\boldsymbol { \Phi}}({\bf{G}},\{{\bf{P}}_k\}_{k=1}^K)]
\end{align} where the weighting matrix ${\bf{W}}$ is a positive semi-definite matrix \cite{Beck06}. Here ${\bf{W}}$ is not restricted to be a diagonal matrix. Given any two matrices ${\boldsymbol{N}}$ and ${\boldsymbol{M}}$ satisfying ${\boldsymbol{N}} \succeq {\boldsymbol M}\succeq {\bf{0}}$, we have ${\rm{Tr}}[{\bf{W}}{\boldsymbol{N}}] \ge {\rm{Tr}}[{\bf{W}}{\boldsymbol{M}}]$. The weighted MSE is thus a monotonically matrix-increasing function of ${\boldsymbol { \Phi}}({\bf{G}},\{{\bf{P}}_k\}_{k=1}^K)$ \cite{Jorswieck07}.
Clearly, transceiver design with weighted MSE minimization aims at minimizing the distortion between the recovered and the transmitted signal \cite{Sampth01,Palomar03,Guan08}.

(2) \textit{Capacity}: Capacity maximization is another important and widely used performance metric for transceiver design. Denoting the received pilot for channel estimation as ${\bf{r}}$, the channel capacity between the source and destination is $\textsl{I}({\bf{s}};{\bf{y}}|{\bf{r}})$ \cite{Hassibi03}. To the best of our knowledge, the exact capacity of MIMO channels with channel estimation errors is still open even for point-to-point MIMO systems \cite{Ding10,Hassibi03}. However, a lower bound of the capacity can be found as
\begin{align}
\label{bound_capacity}
 -{\rm{log}}|{\boldsymbol {\Phi}}({\bf{G}},\{{\bf{P}}_k\}_{k=1}^K) | \le \textsl{I}({\bf{s}};{\bf{y}}|{\bf{r}}).
\end{align}
The equality in (\ref{bound_capacity}) holds when perfect CSI is known \cite{Palomar03,Tang07}. For imperfect CSI, the tightness of this bound is extensively investigated in \cite{Yoo2006,Hassibi03}. This lower bound $-{\rm{log}}|{\boldsymbol {\Phi}}({\bf{G}},\{{\bf{P}}_k\}_{k=1}^K)|$ can be interpreted as the sum-rate of multiple transmitted data streams when linear equalizer ${\bf{G}}$ is employed. It has been widely used to replace the unknown exact capacity as a performance metric. Based on this lower bound, the robust transceiver design maximizing capacity can be replaced by minimizing the following objective function \cite{Yoo2006,Ding10}:
\begin{align}
\label{Obj 2}
\text{Obj 2:}\ {\rm{log}}|{\boldsymbol {\Phi}}({\bf{G}},\{{\bf{P}}_k\}_{k=1}^K)|.
\end{align}

(3) \textit{Worst MSE}: Notice that capacity maximization criterion (\textbf{Obj 2}) does not impose any fairness on the simultaneously transmitted multiple data streams, while  the weighted MSE minimization criterion (\textbf{Obj 1}) imposes only a limited degree of fairness on the data as it involves only a linear operation on the MSE. When fairness is required to balance the performance across different data streams, worst MSE minimization is a good alternative for such transceiver design. In general, the worst MSE can be represented as \cite{Palomar03}
\begin{align}
\label{Obj 3}
\text{Obj 3:}\ &{\boldsymbol \psi}_{1}[{\rm{d}}({\boldsymbol \Phi}({\bf{G}},\{{\bf{P}}_k\}_{k=1}^K))]
\end{align}where ${\boldsymbol \psi}_{1}(\bullet)$ is an increasing Schur-convex function and ${\rm{d}}({\boldsymbol \Phi}({\bf{G}},\{{\bf{P}}_k\}_{k=1}^K))$ denotes a vector consisting of the diagonal elements of ${\boldsymbol \Phi}({\bf{G}},\{{\bf{P}}_k\}_{k=1}^K)$, i.e.,
\begin{align}
{\rm{d}}({\boldsymbol \Phi}({\bf{G}},\{{\bf{P}}_k\}_{k=1}^K))=\left[[{\boldsymbol \Phi}({\bf{G}},\{{\bf{P}}_k\}_{k=1}^K)]_{1,1}\ \cdots \ [{\boldsymbol \Phi}({\bf{G}},\{{\bf{P}}_k\}_{k=1}^K)]_{N,N} \ \right]^{\rm{T}},
\end{align}
with the symbol $[{\bf{Z}}]_{i,j}$ representing the $(i,j)^{\rm{th}}$ entry of ${\bf{Z}}$. It follows that ${\boldsymbol \psi}_{1}[{\rm{d}}({\boldsymbol \Phi}({\bf{G}},\{{\bf{P}}_k\}_{k=1}^K))]$ is also a monotonically  matrix increasing function with respect to ${\boldsymbol { \Phi}}({\bf{G}},\{{\bf{P}}_k\}_{k=1}^K)$. Note that the objective function in (\ref{Obj 3}) is applicable to other design criteria involving fairness considerations.

(4) \textit{Weighted sum rate}: When a preference is required to be given to a certain data stream (e.g., loading more resources to the data streams with better channel state information so that the weighted sum rate is maximized), the objective function can be written as \cite{Palomar03}
\begin{align}
\label{Obj 4}
\text{Obj 4:}\ &{\boldsymbol \psi}_{2}[{\rm{d}}({\boldsymbol \Phi}({\bf{G}},\{{\bf{P}}_k\}_{k=1}^K))]
\end{align}
where ${\boldsymbol \psi}_{2}(\bullet)$ is an increasing Schur-concave function. Similarly to (\ref{Obj 3}), this function ${\boldsymbol \psi}_{2}[{\rm{d}}({\boldsymbol \Phi}({\bf{G}},\{{\bf{P}}_k\}_{k=1}^K))]$ is a monotonically  matrix increasing function with respect to ${\boldsymbol { \Phi}}({\bf{G}},\{{\bf{P}}_k\}_{k=1}^K)$.

\noindent {\textbf{\underline{\textsl{Remark 1:}}}} Some objective functions on signal to inference plus noise ratio (SINR) and bit error rate (BER) can also be formulated as (\ref{Obj 3}) or (\ref{Obj 4}) and thus can be incorporated into our framework. For example, when the objective is to maximize a sum of weighted SINRs, the objective function can be formulated as the form of (\ref{Obj 4}) as an increasing Schur-concave function of the diagonal elements of the MSE matrix. Similarly, when the objective is to maximize the harmonic mean of SINRs or to maximize the minimal SINR, the objective function can be formulated as the form of (\ref{Obj 3}) as an increasing Schur-convex function of the diagonal elements of the MSE matrix. On the other hand, when BER minimization is concerned, when all the data streams are modulated using the same scheme, the average BER can be approximated as an increasing Schur-convex function of the diagonal elements of the MSE matrix \cite{Palomar2006} and can be incorporated into the category of Objective 3 in (\ref{Obj 3}).

\subsection{Problem Formulation}
Although the above four criteria aim at different objectives, they have one common feature, that is, the objective functions are monotonically matrix-increasing functions with respect to the data MSE.  The corresponding transceiver design problems can therefore be unified into a single form:
\begin{align}
\label{Opt_X_0000}
\ & \min_{{\bf{P}}_k,{\bf{G}}} \ \ \  {\boldsymbol f}({\boldsymbol \Phi}({\bf{G}},\{{\bf{P}}_k\}_{k=1}^K)) \nonumber \\
& \  {\rm{s.t.}} \ \ \ \ {\rm{Tr}}({\bf{P}}_k{\bf{R}}_{{\bf{x}}_{k-1}}{\bf{P}}_k^{\rm{H}}) \le P_k, \ \ k=1,\cdots, K
\end{align}where ${\boldsymbol f}(\bullet)$ is a real-value matrix monotonically increasing function with ${\boldsymbol \Phi}({\bf{G}},,\{{\bf{P}}_k\}_{k=1}^K)$ as its argument. Notice that the constraints here are imposed on the powers averaged over the channel estimation errors.

With the definition of the data MSE (\ref{MSE_final}) and by differentiating the trace of the MSE with respect to ${\bf{G}}$ and setting the result to zero, we can easily obtain a linear minimum MSE (LMMSE) equalizer as \cite{Kay93}

\begin{align}
\label{G} {\bf{G}}_{\rm{LMMSE}}& =[\prod_{k=1}^K{\bf{\bar H}}_{k}{\bf{P}}_k]^{\rm{H}}[{\bf{\bar
H}}_{K}{\bf{P}}_K{\bf{R}}_{{\bf{x}}_{K-1}}{\bf{P}}_K^{\rm{H}} {\bf{\bar
H}}_{K}^{\rm{H}}+{ {\rm{Tr}}({\bf{P}}_K{\bf{R}}_{{\bf{x}}_{K-1}} {\bf{P}}_K^{\rm{H}}
{\boldsymbol\Psi} _{K}
   ){\boldsymbol\Sigma} _{K}+{\bf{R}}_{n_K}}]^{-1},
\end{align}
with the following property \cite{Beck06,Palomar03}:
\begin{align}
\label{G_OPT}
{\boldsymbol \Phi}({\bf{G}}_{\rm{LMMSE}},\{{\bf{P}}_k\}_{k=1}^K) \preceq {\boldsymbol \Phi}({\bf{G}},\{{\bf{P}}_k\}_{k=1}^K).
\end{align}
The above equality holds when ${\bf{G}}={\bf{G}}_{\rm{LMMSE}}$. Because ${\boldsymbol f}(\bullet)$ is monotonically  matrix-increasing, it follows easily from (\ref{G_OPT}) that ${\boldsymbol f}({\boldsymbol \Phi}({\bf{G}}_{\rm{LMMSE}},\{{\bf{P}}_k\}_{k=1}^K))\le {\boldsymbol f}({\boldsymbol \Phi}({\bf{G}},\{{\bf{P}}_k\}_{k=1}^K))$. It means that ${\boldsymbol f}({\boldsymbol \Phi}({\bf{G}}_{\rm{LMMSE}},\{{\bf{P}}_k\}_{k=1}^K))$ is a tight lower bound of the objective function in (\ref{Opt_X_0000}). Together with the fact that the equalizer ${\bf{G}}$ is not involved in the constraints in (\ref{Opt_X_0000}), the optimization problem in (\ref{Opt_X_0000}) is equivalent to
\begin{align}
\label{Opt_X_000}
\ & \min_{{\bf{P}}_k} \ \ \  {\boldsymbol f}({\boldsymbol \Phi}({\bf{G}}_{\rm{LMMSE}},\{{\bf{P}}_k\}_{k=1}^K)) \nonumber \\
& \  {\rm{s.t.}} \ \ \ \ {\rm{Tr}}({\bf{P}}_k{\bf{R}}_{{\bf{x}}_{k-1}}{\bf{P}}_k^{\rm{H}}) \le P_k, \ \ k=1,\cdots, K.
\end{align}
It implies that the optimal equalizer of (\ref{Opt_X_0000}) is ${\bf{G}}_{\rm{LMMSE}}$ in (\ref{G}). Substituting the optimal equalizer into ${\boldsymbol \Phi}({\bf{G}},\{{\bf{P}}_k\}_{k=1}^K)$ in (\ref{MSE_final}) and denoting ${\boldsymbol \Phi}({\bf{G}}_{\rm{LMMSE}},\{{\bf{P}}_k\}_{k=1}^K)={\boldsymbol \Phi}_{\rm{MMSE}}(\{{\bf{P}}_k\}_{k=1}^K)$ for simplicity, we have
\begin{align}
\label{MSE_Matrix_Orig}
 {\boldsymbol \Phi}_{\rm{MMSE}}(\{{\bf{P}}_k\}_{k=1}^K)&={\bf{I}}_N-[\prod_{k=1}^K{\bf{\bar H}}_{k}{\bf{P}}_k]^{\rm{H}}[{\bf{\bar
H}}_{K}{\bf{P}}_K{\bf{R}}_{{\bf{x}}_{K-1}}{\bf{P}}_K^{\rm{H}} {\bf{\bar
H}}_{K}^{\rm{H}}\nonumber \\
&+{ {\rm{Tr}}({\bf{P}}_K{\bf{R}}_{{\bf{x}}_{K-1}} {\bf{P}}_K^{\rm{H}}
{\boldsymbol\Psi} _{K}
   ){\boldsymbol\Sigma} _{K}+{\bf{R}}_{n_K}}]^{-1}[\prod_{k=1}^K{\bf{\bar H}}_{k}{\bf{P}}_k].
\end{align}

For multi-hop AF MIMO relaying systems, the received signal at the $k^{\rm{th}}$ relay depends on the forwarding matrices at all preceding relays, causing the power allocations at different relays to be coupled to each other (as seen in the constraints of (\ref{Opt_X_000})), and thus making the problem (\ref{Opt_X_000}) difficult to solve.
To proceed, we define the following new variables in terms of ${\bf{P}}_k$:
\begin{align}
\label{F_definition}
{\bf{F}}_1&\triangleq {\bf{P}}_1{\bf{Q}}_0^{\rm{H}}, \nonumber \\
{\bf{F}}_k&\triangleq
{\bf{P}}_k{\bf{K}}_{{\bf{F}}_{k-1}}^{1/2}(\underbrace{{\bf{K}}_{{\bf{F}}_{k-1}}^{-1/2}{\bf{\bar
H}}_{k-1}{\bf{F}}_{k-1}{\bf{F}}_{k-1}^{\rm{H}}{\bf{\bar
H}}_{k-1}^{\rm{H}}{\bf{K}}_{{\bf{F}}_{k-1}}^{-1/2}+{\bf{I}}_{M_{k-1}}}_{\triangleq {\boldsymbol \Pi}_{k-1}})^{1/2}{\bf{Q}}_{k-1}^{\rm{H}}
\end{align}
where ${\bf{K}}_{{\bf{F}}_k}\triangleq {\rm{Tr}}({\bf{F}}_k{\bf{F}}_k^{\rm{H}}
{\boldsymbol\Psi} _{k}
   ){\boldsymbol\Sigma} _{k}+{\sigma}_{n_k}^2{\bf{I}}_{M_k}$ and ${\bf{Q}}_k$ is an unknown unitary matrix. The introduction of ${\bf{Q}}_k$ is due to the fact
   that for a positive semi-definite matrix ${\bf{M}}$, its square root has the form ${\bf{M}}^{1/2}{\bf{Q}}$ where ${\bf{Q}}$ is a unitary matrix. With the new variables, the MMSE matrix ${\boldsymbol \Phi}_{\rm{MMSE}}(\{{\bf{P}}_k\}_{k=1}^K)$ (\ref{MSE_Matrix_Orig}) is reformulated as
\begin{align}
\label{MSE_Matrix_M}
{\boldsymbol \Phi}_{\rm{MMSE}}(\{{\bf{P}}_k\}_{k=1}^K)
 & ={\bf{I}}_N-{\bf{Q}}_0^{\rm{H}}[\prod_{k=1}^K{\bf{Q}}_k{\boldsymbol \Pi}_k^{-1/2}{\bf{K}}_{{\bf{F}}_k}^{-1/2}{\bf{\bar H}}_k{\bf{F}}_k]^{\rm{H}}[\prod_{k=1}^K{\bf{Q}}_k\underbrace{{\boldsymbol \Pi}_k^{-1/2}{\bf{K}}_{{\bf{F}}_k}^{-1/2}{\bf{\bar H}}_k{\bf{F}}_k}_{\triangleq  {\boldsymbol A}_{k}}]{\bf{Q}}_0\nonumber \\
 &={\bf{I}}_N-{\bf{Q}}_0^{\rm{H}}{\boldsymbol A}_{1}^{\rm{H}}{\bf{Q}}_1^{\rm{H}}\cdots{\boldsymbol A}_{K}^{\rm{H}}{\bf{Q}}_K^{\rm{H}}{\bf{Q}}_K {\boldsymbol A}_{K}\cdots{\bf{Q}}_1{\boldsymbol A}_{1}{\bf{Q}}_0.
\end{align} Meanwhile, the power constraint in the $k^{\rm{th}}$ hop (i.e., ${\rm{Tr}}({\bf{P}}_k{\bf{R}}_{{\bf{x}}_{k-1}}{\bf{P}}_k^{\rm{H}}) \le P_k$) can now be rewritten as
\begin{align}
\label{constraint}
{\rm{Tr}}({\bf{F}}_k{\bf{F}}_k^{\rm{H}}) \le P_k.
\end{align}
It is clear that with the new variables ${\bf{F}}_k$, the constraints become independent of each other. Moreover, the latter transformation of the objective function in the unified problem will not affect the constraints, thus improving the tractability of the problem. Putting (\ref{MSE_Matrix_M}) and (\ref{constraint}) into (\ref{Opt_X_000}), the unified transceiver design problem can be reformulated as
\begin{align}
\label{Opt_X}
{\textbf{P1:}} \ & \min_{{\bf{F}}_k,{\bf{Q}}_k} \ \ \  {\boldsymbol f}( {\bf{I}}_N-{\bf{Q}}_0^{\rm{H}}{\boldsymbol \Theta}{\bf{Q}}_0 ) \nonumber \\
& \  {\rm{s.t.}} \ \ \ \ \ {\rm{Tr}}({\bf{F}}_k{\bf{F}}_k^{\rm{H}}) \le P_k, \ \ k=1,\cdots, K  \nonumber \\
& \ \ \ \ \ \   \  \ \ \  {\boldsymbol \Theta}={\boldsymbol A}_{1}^{\rm{H}}{\bf{Q}}_1^{\rm{H}}\cdots{\boldsymbol A}_{K}^{\rm{H}}{\bf{Q}}_K^{\rm{H}}{\bf{Q}}_K {\boldsymbol A}_{K}\cdots{\bf{Q}}_1{\boldsymbol A}_{1} \nonumber \\
& \ \ \ \ \ \   \ \ \ \ {\bf{Q}}_k^{\rm{H}}{\bf{Q}}_k={\bf{I}}_{M_k}.
\end{align}

From the definition of ${\boldsymbol A}_k$ in (\ref{MSE_Matrix_M}) and noticing that ${\bf{K}}_{{\bf{F}}_k}={\rm{Tr}}({\bf{F}}_k{\bf{F}}_k^{\rm{H}}{\boldsymbol \Psi}_k){\boldsymbol \Sigma}_k+\sigma_{n_k}^2{\bf{I}}_{M_k}$, it can be seen that the design variable ${{\bf{F}}_k}$ appears at multiple positions in the objective function and is involved in matrix inversion and square root operations through ${\bf{K}}_{{\bf{F}}_k}$. This is significantly different from transceiver design for multi-hop MIMO relaying systems with perfect CSI in \cite{Rong2009TWC}. Therefore, the optimization problem is much more complicated than its counterpart with perfect CSI. Indeed, as demonstrated by, e.g., \cite{Xing1012} and \cite{Ding09,Zhang08,Ding10}, robust transceiver design is much more complicated and challenging than its counterpart with perfect CSI even for point-to-point or dual-hop relaying MIMO systems.

%

\section{Optimal Solution for the Robust Transceiver}
\label{sect:structure}

Clearly from the formulation of \textbf{P1} in (\ref{Opt_X}), two sets of matrix variables (i.e., ${\bf{F}}_k,{\bf{Q}}_k$) need to be determined. In this section, their optimal structures will be derived first, which enables the simplification of the optimization problem in (\ref{Opt_X}) into a problem with only scalar variables. An iterative water-filling algorithm is then applied to solve the simplified problem. The relationship between our proposed solution and a number of existing solutions will also be discussed in detail.

\subsection{Optimal ${\bf{Q}}_k$}

Based on the formulations of the objectives given in (\ref{Obj 1}), (\ref{Obj 2}), (\ref{Obj 3}) and (\ref{Obj 4}), we have the following property of the optimization problem  \textbf{P1}.

\noindent \underline{\textsl{\textbf{Property 1:}}} At the optimal value of \textbf{P1}, ${\bf{Q}}_0^{\rm{H}} {\boldsymbol{\Theta}}{\bf{Q}}_0 $  and the objective function ${\boldsymbol f}( {\bf{I}}_N-{\bf{Q}}_0^{\rm{H}}{\boldsymbol \Theta}{\bf{Q}}_0 )$ can be written respectively as
\begin{align}
\label{Q_0}
{\bf{Q}}_0^{\rm{H}} {\boldsymbol \Theta}{\bf{Q}}_0={\bf{U}}_{\boldsymbol \Omega}{\rm{diag}}[{\boldsymbol \lambda}({\boldsymbol \Theta})]{\bf{U}}_{\boldsymbol \Omega}^{\rm{H}},
\end{align}
\begin{align}
{\boldsymbol f}( {\bf{I}}_N-{\bf{Q}}_0^{\rm{H}}{\boldsymbol \Theta}{\bf{Q}}_0 )={\boldsymbol f}({\bf{I}}_N-{\bf{U}}_{\boldsymbol \Omega}{\rm{diag}}[{\boldsymbol \lambda}({\boldsymbol \Theta})]{\bf{U}}_{\boldsymbol \Omega}^{\rm{H}}) \triangleq {\boldsymbol g}[{\boldsymbol \lambda}({\boldsymbol \Theta})],
\end{align}
where ${\boldsymbol  g}(\bullet)$ is a monotonically decreasing and Schur-concave function with respect to ${\boldsymbol \lambda}({\boldsymbol \Theta})$ \footnote{The specific expressions for ${\boldsymbol g}(\bullet)$ are given in Appendix~\ref{Appedix:2}, but they are not important for the derivation of the optimal structures. }; the vector
${\boldsymbol \lambda}({\boldsymbol \Theta})=[{\lambda}_1({\boldsymbol \Theta}), \cdots, {\lambda}_N({\boldsymbol \Theta})]^{\rm{T}}$ with ${\lambda}_i({\boldsymbol \Theta})$ being the $i^{\rm{th}}$ largest eigenvalue of ${\boldsymbol \Theta}$; and
\begin{align}
\label{U}
{\bf{U}}_{\boldsymbol{\Omega}}=\left\{ {\begin{array}{*{20}c}
   {{\bf{U}}_{\bf{W}} \ \ \ \text{for Obj 1}}  \\
   {{\bf{U}}_{\rm{Arb}}\ \ \text{for Obj 2}}  \\
   {{\bf{Q}}_{\rm{DFT}} \ \ \text{for Obj 3}}  \\
   { \ \ {\bf{I}}_N \ \ \ \  \text{for Obj 4}}  \\
\end{array}} \right ..
\end{align}
In (\ref{U}), the matrix ${\bf{U}}_{\bf{W}}$ is unitary and defined from the eigen-decomposition of the weighting matrix ${\bf{W}}$, i.e., ${\bf{W}}={\bf{U}}_{\bf{W}}{\boldsymbol{\Lambda}}_{\bf{W}}{\bf{U}}_{\bf{W}}^{\rm{H}}$ with ${\boldsymbol{\Lambda}}_{\bf{W}}\searrow$; the matrix ${\bf{U}}_{\rm{Arb}}$ is an arbitrary unitary matrix; and ${\bf{Q}}_{\rm{DFT}}$ is the discrete Fourier transform (DFT) matrix making ${\bf{Q}}_{\rm{DFT}}{\rm{diag}}[{\boldsymbol \lambda}({\boldsymbol \Theta})]{\bf{Q}}_{\rm{DFT}}^{\rm{H}}$ have identical diagonal elements.

\noindent \textbf{Proof:} See Appendix~\ref{Appedix:2}. $\blacksquare$

We notice that the equality in (\ref{Q_0}) will hold directly, when
\begin{align}
\label{Q_0_opt}
 {\bf{Q}}_0={\bf{U}}_{{\boldsymbol \Theta}}{\bf{U}}_{\boldsymbol{\Omega}}^{\rm{H}}
\end{align}
where ${\bf{U}}_{\boldsymbol{\Theta}}$ is the unitary matrix corresponding to the eigen-decomposition of ${\boldsymbol{\Theta}}$ with eigenvalues in decreasing order. Since ${\bf{Q}}_0$ is not involved in the constraints in (\ref{Opt_X}), it follows from \textbf{Property 1} that ${\bf{Q}}_0={\bf{U}}_{{\boldsymbol \Theta}}{\bf{U}}_{\boldsymbol{\Omega}}^{\rm{H}}$ is the optimal solution of ${\bf{Q}}_0$ for \textbf{P1}.

Using \textbf{Property 1}, the objective function of (\ref{Opt_X}) can be directly replaced by ${\boldsymbol g} [{\boldsymbol \lambda}({\boldsymbol \Theta})]$ and thus the optimization problem is equivalent to
\begin{align}
\label{P_1_2}
{\textbf{P2:}} \  & \min_{{\bf{F}}_k,{\bf{Q}}_k} \ \ \  {\boldsymbol g} [{\boldsymbol \lambda}({\boldsymbol \Theta})] \nonumber \\
& \ \ {\rm{s.t.}} \ \ \ \  {\boldsymbol \Theta}={\boldsymbol A}_{1}^{\rm{H}}{\bf{Q}}_1^{\rm{H}}\cdots{\boldsymbol A}_{K}^{\rm{H}}{\bf{Q}}_K^{\rm{H}}{\bf{Q}}_K {\boldsymbol A}_{K}\cdots{\bf{Q}}_1{\boldsymbol A}_{1} \nonumber \\
& \ \  \ \ \ \ \ \ \ \ {\rm{Tr}}({\bf{F}}_k{\bf{F}}_k^{\rm{H}}) \le P_k, \ \ {\bf{Q}}_k^{\rm{H}}{\bf{Q}}_{k}={\bf{I}}_{M_k}, \ \  k=1,\cdots,K.
\end{align} For this optimization problem, we have another property as follows.

\noindent \underline{\textsl{\textbf{Property 2:}}} As ${\boldsymbol g}(\bullet)$ is a decreasing and Schur-concave function, the objective function in \textbf{P2} satisfies
\begin{align}
&{\boldsymbol g}({\boldsymbol \lambda}({\boldsymbol \Theta}))\ge{\boldsymbol g}( \underbrace{[{\gamma}_{1}(\{{\bf{F}}_k\}_{k=1}^K) \ \cdots {\gamma}_{N}(\{{\bf{F}}_k\}_{k=1}^K)]^{\rm{T}}}_{\triangleq {\boldsymbol {\gamma}}(\{{\bf{F}}_k\}_{k=1}^K)} ) \label{inequ_3} \\
\text{with} \ \ & {\gamma}_{i}(\{{\bf{F}}_k\}_{k=1}^K)\triangleq \prod_{k=1}^K\frac{{ \lambda}_i({\bf{ F}}_k^{\rm{H}}{\bf{\bar H}}_k^{\rm{H}}{\bf{K}}_{{\bf{F}}_k}^{-1}{\bf{\bar{H}}}_k
 {\bf{ F}}_k)}{1+{ \lambda}_i({\bf{ F}}_k^{\rm{H}}{\bf{\bar H}}_k^{\rm{H}}{\bf{K}}_{{\bf{F}}_k}^{-1}{\bf{\bar{H}}}_k
 {\bf{ F}}_k)} \label{Property2_gamma},
\end{align}with the equality in (\ref{inequ_3}) holds when
\begin{align}
\label{Q_k_opt}
{\bf{Q}}_k={\bf{V}}_{{\boldsymbol{A}}_{k+1}}{\bf{U}}_{{\boldsymbol{A}}_{k}}^{\rm{H}}, \ \ k=1,\cdots,K-1,
\end{align}
and ${\bf{Q}}_K$ is an arbitrary unitary matrix. In (\ref{Q_k_opt}), unitary matrices ${\bf{U}}_{{\boldsymbol{A}}_k}$ and ${\bf{V}}_{{\boldsymbol{A}}_{k}}$ are defined based on the singular value decomposition (SVD) ${\boldsymbol{A}}_k={\bf{U}}_{{\boldsymbol{A}}_k}
{\boldsymbol{\Lambda}}_{{\boldsymbol{A}}_k}{\bf{V}}_{{\boldsymbol{A}}_k}^{\rm{H}} \ \ \text{with} \  \ {\boldsymbol{\Lambda}}_{{\boldsymbol{A}}_k} \searrow$.

\noindent \textbf{Proof:} See Appendix~\ref{Appedix:3}. $\blacksquare$

It is clear from (\ref{Q_k_opt}) that ${\bf{Q}}_k, k=1,\cdots,K-1$ can be uniquely computed from ${\boldsymbol{A}}_k$ which is determined only by ${\bf{F}}_k$ as shown in (\ref{MSE_Matrix_M}). Similarly, according to (\ref{Q_0_opt}) and the definition of ${\boldsymbol{\Theta}}$, it can be concluded that ${\bf{Q}}_0$ is determined by ${\bf{Q}}_k, k=1,\cdots,K$ and ${\boldsymbol{A}}_k$, and therefore it is eventually determined only by ${\bf{F}}_k$. With this fact and \textbf{Property 2}, the optimization problem with two set of variables of ${\bf{F}}_k$ and ${\bf{Q}}_k$ in \textbf{P2} (\ref{P_1_2}) can be reduced to the optimization problem with only one set of variables of ${\bf{F}}_k$ as follows:
\begin{align}
\label{Opt_X_FinalB}
{\textbf{P3:}} \ & \min_{{\bf{F}}_k} \ \ \  {\boldsymbol g} [{\boldsymbol \gamma}(\{{\bf{F}}_k\}_{k=1}^K)]\nonumber \\
& \ {\rm{s.t.}} \ \ \ \  {\gamma}_{i}(\{{\bf{F}}_k\}_{k=1}^K)=\prod_{k=1}^K\frac{{ \lambda}_i({\bf{ F}}_k^{\rm{H}}{\bf{\bar H}}_k^{\rm{H}}{\bf{K}}_{{\bf{F}}_k}^{-1}{\bf{\bar{H}}}_k
 {\bf{ F}}_k)}{1+{ \lambda}_i({\bf{ F}}_k^{\rm{H}}{\bf{\bar H}}_k^{\rm{H}}{\bf{K}}_{{\bf{F}}_k}^{-1}{\bf{\bar{H}}}_k
 {\bf{ F}}_k)} \nonumber \\
&  \ \ \ \ \ \ \ \ \ \ {\rm{Tr}}({\bf{F}}_k{\bf{F}}_k^{\rm{H}}) \le P_k, \ \  k=1,\cdots,K
\end{align}

\subsection{Optimal Structure of ${\bf{F}}_k$}

Since ${\boldsymbol g}(\bullet)$ is a monotonically decreasing function of its vector argument, we have the following additional property of the optimal solution of ${\bf{F}}_k$ in \textbf{P3}.


\noindent \textbf{\underline{\textsl{Property 3:}}} The optimal solutions of the optimization problem \textbf{P3} in (\ref{Opt_X_FinalB}) always occur on the boundary, i.e., ${\rm{Tr}}({\bf{F}}_{k}{\bf{F}}_{k}^{\rm{H}})=P_k$ and the power constraint is equivalent to
\begin{align}
{\rm{Tr}}[{\bf{F}}_k
{\bf{F}}_k^{\rm{H}}(\alpha_k P_k{\boldsymbol \Psi}_k+\sigma_{n_k}^2{\bf{I}}_{N_k})]/\eta_{f_k}=P_k,
\end{align}
where $\alpha_k$ is a constant as $\alpha_k={\rm{Tr}}({\boldsymbol \Sigma}_k)/M_k$ and
\begin{align}
\label{eta_f}
\eta_{f_k} \triangleq {\rm{Tr}}({\bf{F}}_k{\bf{F}}_k^{\rm{H}}{\boldsymbol \Psi}_k)\alpha_k+\sigma_{n_k}^2.
 \end{align}

\noindent \textbf{Proof:} See Appendix~\ref{Appedix:4}. $\blacksquare$

With \textbf{Property 3}, the optimal solution of the optimization problem (\ref{Opt_X_FinalB}) is exactly the optimal solution of the following optimization problem with different constraints:
\begin{align}
\label{Opt_X_Final}
{\textbf{P4:}} \ & \min_{{\bf{F}}_k} \ \ \  {\boldsymbol g} [{\boldsymbol \gamma}(\{{\bf{F}}_k\}_{k=1}^K)]\nonumber \\
& \ {\rm{s.t.}} \ \ \ \  {\gamma}_{i}(\{{\bf{F}}_k\}_{k=1}^K)=\prod_{k=1}^K\frac{{ \lambda}_i({\bf{ F}}_k^{\rm{H}}{\bf{\bar H}}_k^{\rm{H}}{\bf{K}}_{{\bf{F}}_k}^{-1}{\bf{\bar{H}}}_k
 {\bf{ F}}_k)}{1+{ \lambda}_i({\bf{ F}}_k^{\rm{H}}{\bf{\bar H}}_k^{\rm{H}}{\bf{K}}_{{\bf{F}}_k}^{-1}{\bf{\bar{H}}}_k
 {\bf{ F}}_k)} \nonumber \\
&  \ \ \ \ \ \ \ \ \ {\rm{Tr}}[{\bf{F}}_k
{\bf{F}}_k^{\rm{H}}(\alpha_k P_k{\boldsymbol \Psi}_k+\sigma_{n_k}^2{\bf{I}}_{N_k})]/\eta_{f_k}=P_k.
\end{align}

Now defining unitary matrices ${\bf{U}}_{{\boldsymbol{\mathcal{H}}}_k}$ and ${\bf{V}}_{{\boldsymbol{\mathcal{H}}}_k}$ based on the following SVD:
\begin{align}
\label{decomp}
& ({\bf{K}}_{{\bf{F}}_k}/\eta_{f_k})^{-1/2}{\bf{\bar H}}_k(\alpha_k P_k{\boldsymbol \Psi}_k+{\sigma}_{n_k}^2{\bf{I}}_{N_k})^{-1/2}={\bf{U}}_{{\boldsymbol{\mathcal{H}}}_k}{\boldsymbol \Lambda}_{{\boldsymbol{\mathcal{H}}}_k}{\bf{V}}_{{\boldsymbol{\mathcal{H}}}_k}^{\rm{H}}
\end{align}
with singular values in decreasing order, we have the key result about the optimal structure of ${\bf{F}}_k$ as follows.

\noindent \textbf{\underline{\textsl{Property 4:}}} When ${\boldsymbol{ \Psi}}_k \propto {\bf{I}}$ or ${\boldsymbol \Sigma}_k \propto {\bf{I}}$, the matrix ${\bf{K}}_{{\bf{F}}_k}/\eta_{f_k}$ is constant and independent of ${\bf{F}}_k$. Meanwhile, the optimal solution of the optimization problem (\ref{Opt_X_Final}) has the following structure:
\begin{align}
\label{Structure_F}
& {\bf{F}}_{k,\rm{opt}}=\sqrt{{\xi}_k({\boldsymbol { \Lambda}}_{{\boldsymbol{\mathcal{F}}}_k})}(\alpha_k P_k{\boldsymbol
\Psi}_k+\sigma_{n_k}^2{\bf{I}}_{N_k})^{-1/2}{\bf{V}}_{{\boldsymbol{\mathcal{H}}}_k,N}
{\boldsymbol { \Lambda}}_{{\boldsymbol{\mathcal{F}}}_k}{\bf{U}}_{{\boldsymbol{\mathcal{F}}}_k,N}^{\rm{H}},
\end{align}
where ${\bf{V}}_{{\boldsymbol{\mathcal{H}}}_k,N}$ and ${\bf{U}}_{{\boldsymbol{\mathcal{F}}}_k,N}$ are the matrices consisting of the first $N$ columns of  ${\bf{V}}_{{\boldsymbol{\mathcal{H}}}_k}$ and ${\bf{U}}_{{\boldsymbol{\mathcal{F}}}_k}$, respectively; ${\bf{U}}_{{\boldsymbol{\mathcal{F}}}_k}$ is an arbitrary unitary matrix; ${\boldsymbol{ \Lambda}}_{{\boldsymbol{\mathcal{F}}}_k}$ is a $N \times N$ unknown diagonal matrix; and the scalar ${ \xi}_k({\boldsymbol { \Lambda}}_{{\boldsymbol{\mathcal{F}}}_k})$ is a function of ${\boldsymbol{ \Lambda}}_{{\boldsymbol{\mathcal{F}}}_k}$ and equals
\begin{align}
{ \xi}_k({\boldsymbol { \Lambda}}_{{\boldsymbol{\mathcal{F}}}_k})&={\sigma_{n_k}^2}/\{1-\alpha_k {\rm{Tr}}[{\bf{V}}_{{\boldsymbol{\mathcal{H}}}_k,N}^{\rm{H}}(\alpha_k P_k{\boldsymbol
\Psi}_k+\sigma_{n_k}^2{\bf{I}}_{N_k})^{-1/2}{\boldsymbol
\Psi}_k(\alpha_k P_k{\boldsymbol
\Psi}_k+\sigma_{n_k}^2{\bf{I}}_{N_k})^{-1/2}
{\bf{V}}_{{\boldsymbol{\mathcal{H}}}_k,N}{\boldsymbol{ \Lambda}}_{{\boldsymbol{\mathcal{F}}}_k}^2]\}\nonumber \\
&=\eta_{f_k}.
\end{align}

\noindent \textbf{{Proof:}} See Appendix~\ref{Appedix:5}. $\blacksquare$

\noindent {\textbf{\underline{\textsl{Remark 2:}}}} When reversing the direction of data transmission in the multi-hop system, we can get a dual multi-hop system where the estimated channel matrix in its $(K-k+1)^{\rm{th}}$ hop becomes ${\bf{\bar H}}_k^{\rm{H}}$ and the roles of row correlation matrices and column correlation matrices are interchanged. Using (\ref{F_definition}) and \textbf{Property 4} and after some tedious manipulation, the optimal precoder matrices ${\bf{P}'}_{k,{\rm{opt}}}$ for the dual multi-hop system can be found to be $\beta_{k}{\bf{P}}_{k,{\rm{opt}}}^{\rm{H}}$ where $\beta_{k}$ is a scalar. This means that there exists an uplink-downlink duality in the multi-hop AF MIMO relaying systems with channel estimation errors.


In the optimal structure given by (\ref{Structure_F}), the scalar
variable ${ \xi}_k({\boldsymbol { \Lambda}}_{{\boldsymbol{\mathcal{F}}}_k})$
can be uniquely determined by the matrix ${\boldsymbol { \Lambda}}_{{\boldsymbol{\mathcal{F}}}_k}$
and therefore the only unknown variable in (\ref{Structure_F})
is ${\boldsymbol { \Lambda}}_{{\boldsymbol{\mathcal{F}}}_k}$. The computation of ${\boldsymbol { \Lambda}}_{{\boldsymbol{\mathcal{F}}}_k}$
will be addressed in detail in the following subsection.

\subsection{Computation of ${\boldsymbol { \Lambda}}_{{\boldsymbol{\mathcal{F}}}_k}$}
\label{sect:application}

Substituting the optimal structures given by \textbf{Property 4} into \textbf{P4} and defining
$[{\boldsymbol {\Lambda}}_{{\boldsymbol{\mathcal{H}}}_k}]_{i,i}={h}_{k,i}$ and
$[{\boldsymbol { \Lambda}}_{{\boldsymbol{\mathcal{F}}}_k}]_{i,i}=f_{k,i}$ for $i=1,\cdots, N$, the optimization problem for computing ${\boldsymbol { \Lambda}}_{{\boldsymbol{\mathcal{F}}}_k}$ becomes
\begin{align}
\label{Opt_F}
 & \min_{{f}_{k,i}} \ \ \  {\boldsymbol g} [{\boldsymbol \gamma}(\{{\bf{F}}_k\}_{k=1}^K)]\nonumber \\
& \ {\rm{s.t.}} \ \ \ \ { \gamma}_{i}(\{{\bf{F}}_k\}_{k=1}^K)=\frac{{\prod_{k=1}^K}f_{k,i}^2h_{k,i}^2}{\prod_{k=1}^K(f_{k,i}^2h_{k,i}^2+1)}
\nonumber \\ &  \ \ \ \ \ \ \ \ \ \sum_{i=1}^N f_{k,i}^2=P_k.
\end{align}
The exact expression for ${\boldsymbol g} (\bullet)$ depends on the specific design criterion used for transceiver design. For all four criteria discussed in Section \ref{sect:objectives}, a widely used and computationally efficient iterative algorithm \cite{Yu04} can be applied to solve for $f_{k,i}$ from (\ref{Opt_F}), although the optimization problem (\ref{Opt_F}) is non-convex in nature \cite{Zhang2011}. For completeness, the optimal solution for $f_{k,i}$ will be given case by case in the following.


\subsubsection{Weighted MSE Minimization}

For weighted MSE minimization, it is proved in Appendix~\ref{Appedix:2} that ${\boldsymbol g} [{\boldsymbol \gamma}(\{{\bf{F}}_k\}_{k=1}^K)]=\sum_{i=1}^N(w_i-w_i{\gamma}_{i}(\{{\bf{F}}_k\}_{k=1}^K ))$ where $w_i=[{\boldsymbol \Lambda}_{\bf{W}}]_{i,i}$. Therefore, the optimization problem (\ref{Opt_F}) can be rewritten as
\begin{align}
\label{opt_scalar_MMSE}
&\min_{f_{k,i}} \ \sum_{i=1}^N\left(w_i-\frac{w_i\prod_{k=1}^K f_{k,i}^2h_{k,i}^2}
{\prod_{k=1}^K (f_{k,i}^2h_{k,i}^2+1)}\right) \nonumber \\
 &\ {\rm{s.t.}} \ \   \sum_{i=1}^N f_{k,i}^2 =P_k.
\end{align}
Using the iterative water-filling algorithm, $f_{k,i}$ can be directly computed with given $f_{l,i}$'s where $l\not=k$ as
\begin{align}
\label{f_diag}
{f}_{k,i}^2=\left(\sqrt{\frac{w_i}{\mu_kh_{k,i}^2 }} \sqrt{\prod_{l\not=k}\{{\frac{f_{l,i}^2h_{l,i}^2}{1+f_{l,i}^2
h_{l,i}^2}}\}}
-\frac{1}{h_{k,i}^2}\right)^{+}, \ \ i=1,\cdots,N
,
\end{align}
where $\mu_k$ is the Lagrange multiplier that makes $\sum_{i=1}^N f_{k,i}^2=P_k$. Notice that this iterative water-filling algorithm is guaranteed to converge, as discussed in \cite{Yu04}.

\subsubsection{Capacity Maximization}

As proved in Appendix~\ref{Appedix:2}, the objective function for capacity maximization is given by ${\boldsymbol g}[{\boldsymbol \gamma}(\{{\bf{F}}_k\}_{k=1}^K)]=\sum_{i=1}^{N}{\rm{log}} \left(1- {\gamma}_{i}(\{{\bf{F}}_k\}_{k=1}^K)\right)$, based on which the optimization problem (\ref{Opt_F}) can be written as
\begin{align}
\label{opt:capacity_b}
& \min_{f_{k,i}} \ \ \ \sum_{i=1}^{N}{\rm{log}} \left(1- \frac{\prod_{k=1}^Kf_{k,i}^2h_{k,i}^2}{\prod_{k=1}^K(f_{k,i}^2h_{k,i}^2+1)}\right) \nonumber \\
& \ {\rm{s.t.}} \ \ \ \ \sum_{i=1}^N f_{k,i}^2 =P_k.
\end{align}
Similarly, the iterative water-filling algorithm can be used to solve for $f_{k,i}$ with guaranteed convergence. More specifically, when the $f_{l,i}$'s are given with $l\not=k$, the solution for $f_{k,i}$ can be derived as
\begin{align}
\label{solution_capacity}
&f_{k,i}^2=\frac{1}{h_{k,i}^2}\left( \frac{-a_{k,i}+
\sqrt{a_{k,i}^2+4(1-a_{k,i})a_{k,i}h_{k,i}^2/\mu_k}}{2(1-a_{k,i})}-1 \right)^{+} \ \ i=1,\cdots,N \nonumber\\
& \text{with} \ \ a_{k,i}=\prod_{l\not=k}f_{l,i}^2h_{l,i}^2/(f_{l,i}^2h_{l,i}^2+1)
\end{align}where $\mu_k$ is the Lagrange multiplier that makes $\sum_{i=1}^N f_{k,i}^2 =P_k$ hold.

\subsubsection{MAX-MSE Minimization}

MAX-MSE is in fact a special case of \textbf{Obj 3} and in this case,
${\boldsymbol \psi}_1({\rm{d}}({\boldsymbol \Phi}_{\rm{MSE}}(\{{\bf{P}}_k\}_{k=1}^K)))=\max [{\boldsymbol \Phi}_{\rm{MSE}}(\{{\bf{P}}_k\}_{k=1}^K)]_{i,i}$. As shown in Appendix~\ref{Appedix:2}, ${\boldsymbol g} ({\boldsymbol \lambda}(\{{\bf{F}}_k\}_{k=1}^K))={\boldsymbol \psi}_1 [{\bf{1}}_N-({\sum}_{i=1}^N{\lambda}_{i}(\{{\bf{F}}_k\}_{k=1}^K)/N )\otimes{\bf{1}}_N]$. It follows that ${\boldsymbol g} [{\boldsymbol \gamma}(\{{\bf{F}}_k\}_{k=1}^K)]$ equals
\begin{align}
{\boldsymbol g}[{\boldsymbol \gamma}(\{{\bf{F}}_k\}_{k=1}^K)]&=\max \left\{{\bf{1}}_N-({\sum}_{i=1}^N{\gamma}_{i}(\{{\bf{F}}_k\}_{k=1}^K)/N )\otimes{\bf{1}}_N \right\}\nonumber \\
&= 1-\frac{1}{N}\sum_{i=1}^N{\gamma}_{i}(\{{\bf{F}}_k\}_{k=1}^K).
\end{align}
Clearly this expression for ${\boldsymbol g} (\bullet)$ is similar to that for weighted MSE minimization. The optimal solution for $f_{k,i}$ can then be easily found as (\ref{f_diag}) with $w_i=1$.

\subsubsection{Weighted Sum-Rate Maximization}
Under weighted sum-rate maximization, the objective function \textbf{Obj 4} can be further specified as
\[
{\boldsymbol \psi}_2({\rm{d}}({\boldsymbol \Phi}_{\rm{MMSE}}(\{{\bf{P}}_k\}_{k=1}^K)))=\sum_{i=1}^N v_i{\rm{log}}({\rm{d}}_{[N-i+1]}({\boldsymbol \Phi}_{\rm{MMSE}}(\{{\bf{P}}_k\}_{k=1}^K)))\]
where $v_i$ is the $i^{\rm{th}}$ largest positive weighting factor and ${\rm{d}}_{[i]}({\boldsymbol \Phi}_{\rm{MMSE}}(\{{\bf{P}}_k\}_{k=1}^K))$ is the $i^{\rm{th}}$ largest diagonal element. Roughly speaking, this design scheme exhibits preference for data streams with better channel state information. It is proved in Appendix~\ref{Appedix:2} that for this objective, ${\boldsymbol g} ({\boldsymbol \lambda}(\{{\bf{F}}_k\}_{k=1}^K))={\boldsymbol \psi}_2 [{\bf{1}}_N-{\boldsymbol \lambda}(\{{\bf{F}}_k\}_{k=1}^K)]$. It follows that
\begin{align}
{\boldsymbol g} [{\boldsymbol \gamma}(\{{\bf{F}}_k\}_{k=1}^K)]
&=\sum_{i=1}^{N}v_i{\rm{log}} \left(1- {\gamma}_{i}(\{{\bf{F}}_k\}_{k=1}^K)\right)
\end{align}and the optimization problem is formulated as
\begin{align}
\label{46}
& \min_{f_{k,i}} \ \ \ \sum_{i=1}^{N}v_i{\rm{log}} \left(1- \frac{\prod_{k=1}^Kf_{k,i}^2h_{k,i}^2}{\prod_{k=1}^K(f_{k,i}^2h_{k,i}^2+1)}\right) \nonumber \\
& \ {\rm{s.t.}} \ \ \ \ \sum_{i=1}^N f_{k,i}^2 =P_k.
\end{align}
The optimization problem in (\ref{46}) has a similar form to that in (\ref{opt:capacity_b}), except that there are a number of weighting factors ${v_i}$ in the objective function of (\ref{46}). Therefore, the iterative water-filling solution of $f_{k,i}$ can be obtained similarly to that in (\ref{solution_capacity}) but with $\mu_k$ replaced by $\mu_k/v_k$.

\subsection{Relationship with Existing Solutions}

By comparing our proposed optimal solution given by \textbf{Property 4} with existing solutions for various systems in the literature, we find that our proposed solution reduces to the following existing solutions by setting some system parameters accordingly:

\noindent $\bullet$ the robust design with weighted MSE minimization for a dual-hop AF MIMO relaying system without source precoder in \cite{Xing1012}, by setting $K=2$, ${\boldsymbol \Sigma}_2 \propto {\bf{I}}_{M_2}$, and ${\bf{P}}_1={\bf{I}}_{N}$;

\noindent $\bullet$ the robust design for a dual-hop AF MIMO relaying system in \cite{Xing10}, by setting $K=2$, ${\boldsymbol \Psi}_2 \propto {\bf{I}}_{N_2}$, and ${\bf{P}}_1={\bf{I}}_N$;

%

\noindent $\bullet$ the transceiver design with weighted MSE minimization for a dual hop system with perfect CSI in \cite{Guan08}, by setting $K=2$, ${\boldsymbol \Psi}_k={\boldsymbol \Sigma}_k=\textbf{0}$, ${\bf{W}}={\bf{I}}_N$ and ${\bf{P}}_1={\bf{I}}_N$;

\noindent $\bullet$ the transceiver design for a dual hop system with perfect CSI in \cite{Rong09}, by setting $K=2$, ${\boldsymbol \Psi}_k={\boldsymbol \Sigma}_k=\textbf{0}$ and ${\bf{W}}={\bf{I}}_N$;

\noindent $\bullet$ the transceiver design with capacity maximization for a dual hop system with perfect CSI in \cite{Tang07}, by setting $K=2$, ${\boldsymbol \Psi}_k={\boldsymbol \Sigma}_k=\textbf{0}$ and ${\bf{P}}_1={\bf{I}}_N$;

\noindent $\bullet$ the robust design with weighted MSE minimization for a point-to-point MIMO system in \cite{Ding09}, by setting $K=1$; and

\noindent $\bullet$ the robust design with capacity maximization for a point-to-point MIMO system \cite{Ding10}, by setting $K=1$.

In other words, our proposed solution covers the above designs as special cases. It further verifies the correctness and optimality of our proposed solution.

\subsection{Discussions}

The optimal structure of ${\bf{F}}_k$ in (\ref{Structure_F}) is derived under the condition ${\boldsymbol{ \Psi}}_k \propto {\bf{I}}$ or ${\boldsymbol \Sigma}_k \propto {\bf{I}}$. This condition can be easily satisfied in practice. We notice that the expressions for
${\boldsymbol \Psi}_{k}$ and ${\boldsymbol \Sigma}_{k}$ generally depend on
specific channel estimation algorithms. Denote the transmit
and receive antenna correlation matrices and the channel estimation error variance in the $k^{\rm{th}}$ hop as ${\bf{R}}_{T,k}$, ${\bf{R}}_{R,k}$ and $ \sigma_{e,k}^2$, respectively. Applying the widely used channel estimation algorithms in \cite{Ding09} and \cite{Ding10}, the covariance matrices for channel estimation errors can be written as $ {\boldsymbol{\Psi}}_{k}={\bf{R}}_{T,k}$ and
${\boldsymbol{\Sigma}}_{k}=\sigma_{e,k}^2({\bf{I}}_{M_{k}}+\sigma_{e,k}^2{\bf{R}}_{R,k}^{-1})^{-1}$.
Clearly, when the receive antennas are spaced widely, i.e., ${\bf{R}}_{R,k}\propto {\bf{I}}_{M_{k}}$,  we directly have ${\boldsymbol{\Sigma}}_{k}\propto {\bf{I}}_{M_{k}}$. Moreover, when the length of training is long, the value of $\sigma_{e,k}^2$ will be small and ${\bf{I}}_{M_{k}}+\sigma_{e,k}^2{\bf{R}}_{R,k}^{-1}\approx {\bf{I}}_{M_{k}}$. As a result, ${\boldsymbol{\Sigma}}_{k}$ can be approximated as an identity matrix even when ${\bf{R}}_{R,k}\not\propto {\bf{I}}_{M_{k}}$. Furthermore, when the channel statistics are unknown and the least-squares channel estimator is applied, it can be derived that ${\boldsymbol{\Sigma}}_{k}\propto {\bf{I}}_{M_{k}}$ \cite{Xing1012}. On the other hand, when the transmit antennas are spaced widely, i.e., ${\bf{R}}_{T,k} \propto {\bf{I}}_{N_{k}}$, we can obtain ${\boldsymbol{\Psi}}_{k}\propto {\bf{I}}_{N_{k}}$.

For the general case when  ${\boldsymbol{ \Psi}}_k \not\propto {\bf{I}}_{N_k}$ and ${\boldsymbol \Sigma}_k \not\propto {\bf{I}}_{M_k}$, to the best of our knowledge, finding a closed-form optimal solution of the robust design problem is still open, even for point-to-point MIMO systems \cite{Ding09,Zhang08}. The main difficulty comes from the fact that when ${\boldsymbol{ \Psi}}_k \not\propto {\bf{I}}_{N_k}$ and ${\boldsymbol \Sigma}_k \not\propto {\bf{I}}_{M_k}$, ${\bf{K}}_{{\bf{F}}_k}/\eta_{f_k}$ varies with ${\bf{F}}_k$, and so is not a constant.
However, for this general case, ${\bf{K}}_{{\bf{F}}_k}/\eta_{f_k}$ in (\ref{decomp}) can be replaced by an upper bound
\begin{align}
{\bf{K}}_{{\bf{F}}_k}/\eta_{f_k} \preceq  {P_k\lambda_1({\boldsymbol \Psi}_k)}/({P_k\lambda_1({\boldsymbol \Psi}_k)\alpha_k+\sigma_{n_k}^2}){\boldsymbol \Sigma}_k+{\sigma_{n_k}^2}/({P_k\lambda_1({\boldsymbol \Psi}_k)\alpha_k+\sigma_{n_k}^2}){\bf{I}}_{M_k},
\end{align}such that it is not a function of ${\bf{F}}_k$. Notice that the above equality holds when ${\boldsymbol{ \Psi}}_k \propto {\bf{I}}_{N_k}$ or ${\boldsymbol \Sigma}_k \propto {\bf{I}}_{M_k}$. Then the proposed solution can still be applied for this general case.

When there are two hops ($K=2$), our proposed optimal structure is different from that derived in \cite{Rong20115} (comparing (\ref{Structure_F}) with Equation (16) in \cite{Rong20115}). In \cite{Rong20115}, the solution structure is obtained by implicitly approximating a design-variable-dependent covariance matrix (the matrix A in [14]) as constant. Since the approximation is tight only when the covariance matrix of channel estimation errors seen from the receiver side is proportional to an identity matrix, i.e., ${\boldsymbol \Psi}_k \propto {\bf{I}}_{N_k}, k=1,2$, the proposed solution in \cite{Rong20115} is sub-optimal when ${\boldsymbol \Psi}_k \not\propto {\bf{I}}_{N_k}, k=1,2$. In other words, for dual-hop systems, our proposed solution is optimal under a wider range of cases than that in \cite{Rong20115}, since it is optimal when either ${\boldsymbol \Sigma}_k$ \textbf{or} ${\boldsymbol \Psi}_k$ is proportional to an identity matrix.


With respect to the complexity, it is clear from (\ref{Structure_F}) that the complexity of our algorithm is due to two kinds of operations, i.e., the iterative water-filling computation for the inner diagonal matrix in (\ref{Structure_F}) and the decomposition/multiplication for the matrices on the lefthand and righthand sides of the diagonal matrix in (\ref{Structure_F}). Comparing the structures of the solution in (\ref{Structure_F}) and that in \cite{Rong20115}, similar operations are needed to obtain the solution in [14]. So we can expect that the complexity of our approach is comparable to that in \cite{Rong20115}.

\section{Simulation Results}
\label{sect:simulation}
In this section, the performance of the proposed robust designs is evaluated by simulations. In the simulations, the number of antennas at each node is set to four. At the source node, four independent data streams are transmitted and in each data stream, ${N_{Data}}=10^4$ independent quadrature phase shifting keying (QPSK) symbols
are transmitted. The correlation matrices corresponding to the channel estimation errors are chosen according to the widely used exponential model,
i.e., $[{\boldsymbol{\Psi}}_{k}]_{i,j}=\sigma_e^2\alpha^{|i-j|}$
and  $[{\boldsymbol{\Sigma}}_{k}]_{i,j}=\beta^{|i-j|}$,
where $\alpha$ and $\beta$ are the correlation
coefficients, and $\sigma_{e}^2$ denotes the variance of the channel estimation error \cite{Zhang08,Xing10}. The estimated
channel matrices ${\bf{\bar H}}_{k}$'s, are
generated following the widely used complex Gaussian distributions, ${\bf{\bar H}}_{k}\sim
\mathcal {C}\mathcal
{N}_{M_k,N_k}({\bf{0}}_{M_k,N_k},{(1-\sigma_{e}^2)}/{\sigma_{e}^2}{\boldsymbol
\Sigma}_{k}
\otimes {\boldsymbol \Psi}_{k}^{\rm{T}})$ \cite{Xing10,Musavian07}, such that channel realizations ${\bf{H}}_{k}={\bf{\bar
H}}_{k}+\Delta{\bf{H}}_{k}$ have unit variance. The signal-to-noise ratio (${\rm{SNR}}$) for the $k^{\rm{th}}$ link
is defined as $P_k/\sigma_{n_k}^2$ and each point in the
following figures shows an average result of $10^4$ trials.

\begin{figure}[!ht] \centering
\includegraphics[width=.6\textwidth]{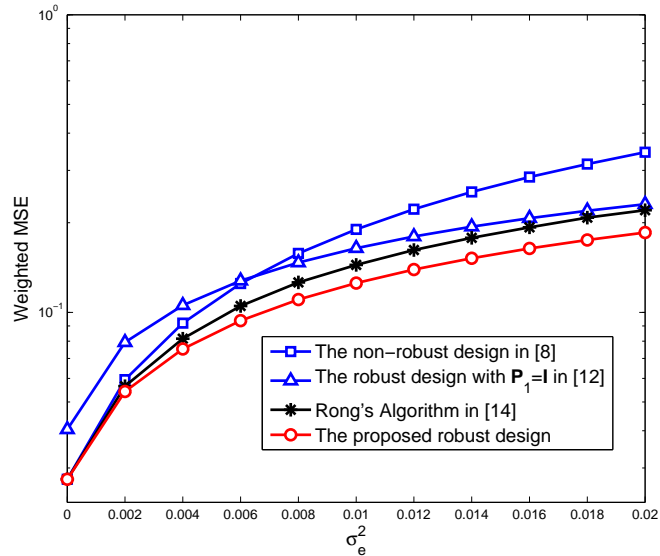}
\caption{Weighted MSE of the detected data in a dual-hop AF relaying system, when $\alpha=0.6$, $\beta=0$ and
$P_k/\sigma_{n_k}^2=30$dB.}\label{fig:2}
\end{figure}

\begin{figure}[!ht]
\centering
\includegraphics[width=.6\textwidth]{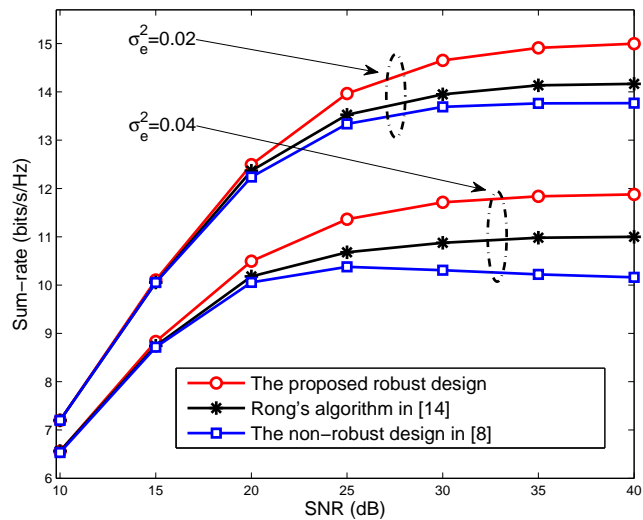}
\caption{Sum-rate of a dual-hop AF relaying system when $\alpha=0.6$ and $\beta=0$.}\label{fig:3}
\end{figure}

A dual hop system ($K=2$) with error correlation coefficients of $\alpha=0.6$ and $\beta=0$ (i.e., ${\boldsymbol \Psi}_k \not\propto {\bf{I}}, {\boldsymbol \Sigma}_k \propto {\bf{I}}, k=1,2$) is considered first. Fig.~\ref{fig:2} shows the weighted MSE at the destination when the weighting matrix is arbitrarily chosen as ${\bf{W}}={\rm{diag}}\{[0.3 \ 0.3 \ 0.26 \ 0.26]\}$ and $P_k/\sigma_{n_k}^2=30$dB. For comparison, the performance of the algorithm based on the estimated channel only (labeled as non-robust design) \cite{Rong09}, the robust algorithm proposed by Rong in \cite{Rong20115} and the robust algorithm without source precoding in \cite{Xing10} is also shown. It is clear from the figure that our proposed robust design offers the best performance, while the non-robust design is the worst. Fig.~\ref{fig:3} shows the sum-rates of various algorithms for the considered two-hop AF MIMO relaying system. It can be seen that the robust algorithms generally have better performance than the algorithm based on estimated CSI only. Furthermore, the performance of the proposed robust design is much better than that of the robust algorithm in \cite{Rong20115}.

\begin{figure}[!ht] \centering
\includegraphics[width=.6\textwidth]{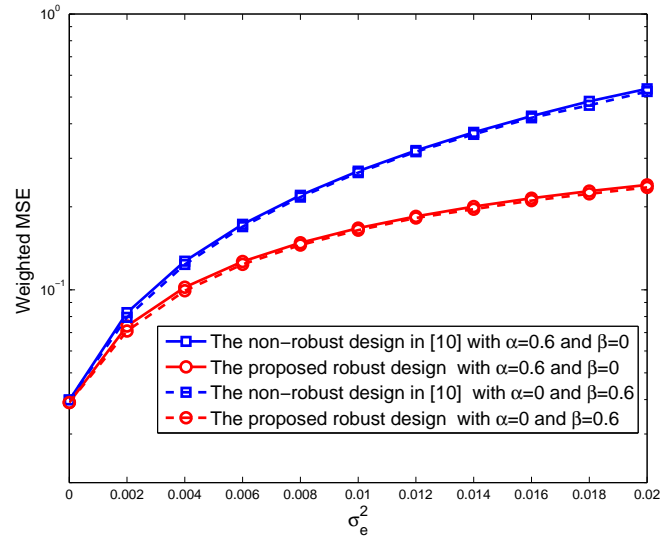}
\caption{Weighted MSE of the detected data in a three-hop AF relaying system when $P_k/\sigma_{n_k}^2=30$dB.}\label{fig:4}
\end{figure}

Next a three-hop AF MIMO relaying system, i.e., $K=3$, is considered to further investigate the effectiveness of the proposed robust design. Since there are few (if any) robust transceiver design algorithms proposed for multi-hop AF MIMO systems in the literature, our proposed robust design is mainly compared with the non-robust design in \cite {Rong2009TWC} in the following. With the weighting matrix being arbitrarily
selected as ${\bf{W}}={\rm{diag}}\{[0.26 \ 0.25 \ 0.25 \ 0.24]\}$,
Fig.~\ref{fig:4} shows the weighted MSE at the destination when $P_k/\sigma_{n_k}^2=30$dB. Here two sets of error correlation coefficients, $(\alpha=0.6, \beta=0)$, and $(\alpha=0, \beta=0.6)$, are taken as examples. They correspond to the cases of $({\boldsymbol \Psi}_k \not\propto {\bf{I}}, {\boldsymbol \Sigma}_k \propto {\bf{I}})$ and $({\boldsymbol \Psi}_k \propto {\bf{I}}, {\boldsymbol \Sigma}_k \not\propto {\bf{I}})$, respectively. It can be seen that the proposed algorithm shows similar performance for the two cases and always outperforms the non-robust design based on the estimated CSI only. When there is no channel estimation error, i.e., $\sigma_e^2=0$, the performance of the two algorithms is the same as expected.

\begin{figure}[!ht] \centering
\includegraphics[width=.6\textwidth]{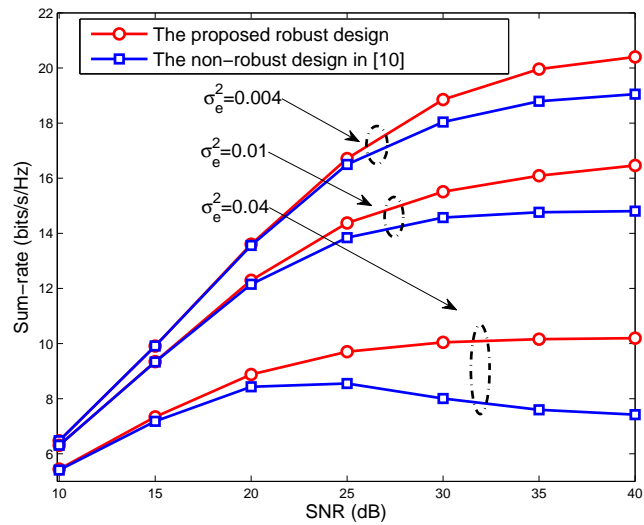}
\caption{Sum-rate of a three-hop AF relaying system when $\alpha=0.6$ and $\beta=0$.}\label{fig:5}
\end{figure}

\begin{figure}[!ht] \centering
\includegraphics[width=.6\textwidth]{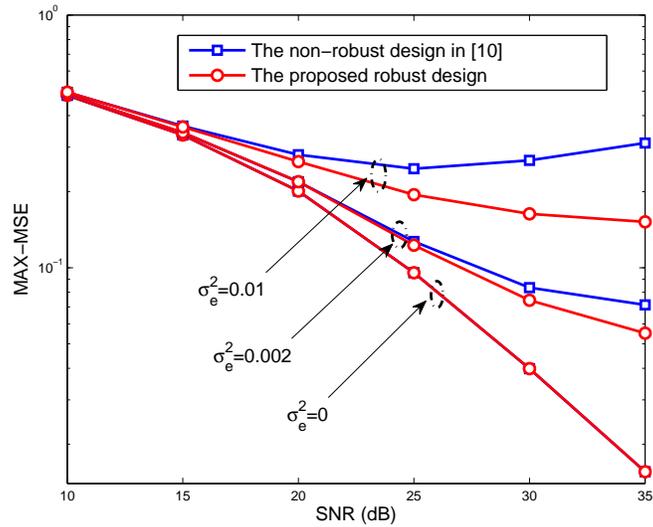}
\caption{Maximum MSE among different received data streams in a three-hop relaying system with $\alpha=0$ and $\beta=0.6$.}\label{fig:6}
\end{figure}

Fig.~\ref{fig:5} shows the sum-rates at different SNRs (${\rm{SNR}}=P_k/\sigma_{n_k}^2$) for the three-hop system. The SNRs at various hops are set as the same for simplicity. The correlation coefficients for the channel estimation errors are taken as $\alpha=0.6$ and $\beta=0$. It is further demonstrated that the proposed algorithm shows better performance than the non-robust algorithm based on estimated CSI only. Furthermore, as the estimation errors increase, the performance gap between the two algorithms enlarges. This result coincides with that for the weighted-MSE-based robust design shown in Fig.~\ref{fig:4}. The performance of the maximum MSE across four data streams with $\alpha=0$ and  $\beta=0.6$ is then shown in Fig.~\ref{fig:6}. Similarly, it is observed that the performance gain of the proposed robust design over the non-robust design with estimated CSI only becomes larger as SNR increases. The performance gap is also more apparent when $\sigma_e^2$ increases.

\begin{figure}[!ht] \centering
\includegraphics[width=.6\textwidth]{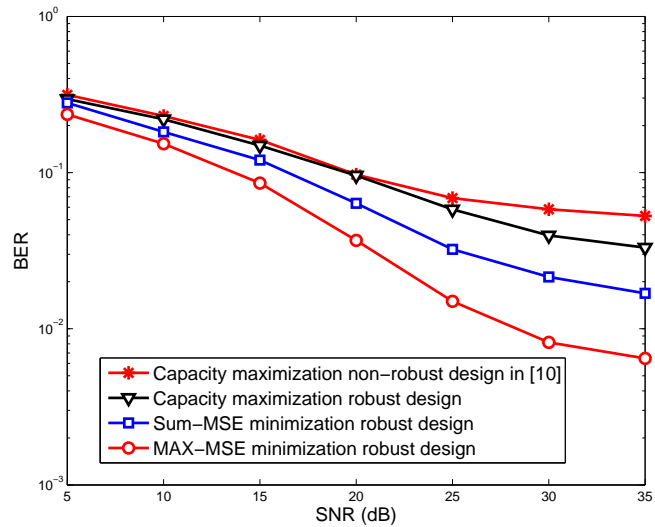}
\caption{BERs of the proposed robust design with different design objectives, when $\alpha=0.6$, $\beta=0$ and $\sigma_e^2=0.004$.}\label{fig:7}
\end{figure}

Finally, Fig.~\ref{fig:7} shows the bit-error-rate (BER) performance for the three-hop systems with different design criteria:  capacity maximization, sum MSE minimization (i.e., weighted MSE minimization with ${\bf{W}}={\bf{I}}$) and MAX-MSE minimization. The parameters are chosen as $\alpha=0.6$, $\beta=0$ and $\sigma_{e}^2=0.004$. It can be seen that in terms of BER performance, the former two criteria perform worse than the latter one since the latter criterion targets the BER performance more. Moreover, the non-robust design with capacity maximization based on estimated CSI only is also given and the results further verify the performance advantage of the proposed robust designs over the non-robust design with estimated CSI only.

\section{Conclusions}
\label{sect:conclusion}
Bayesian robust transceiver design for multi-hop AF MIMO relaying systems with channel estimation errors has been considered.
 Various transceiver design criteria including weighted MSE minimization, capacity maximization, worst MSE minimization and weighted sum-rate maximization have been discussed and formulated into a unified optimization problem. Using majorization theory and properties of matrix-variate functions, the optimal structure of the robust transceivers has been derived. Then the transceiver design problems have been greatly simplified and solved by iterative water-filling algorithm. The performance of the proposed transceiver designs has been demonstrated via simulation results.

\appendices

\section{Proof of Property 1}
\label{Appedix:2}


The proof of Property 1 depends on the specific objective function in (\ref{Opt_X}). In the following, we will discuss the optimization problem (\ref{Opt_X}) with different objective functions case by case.

\noindent \underline{\textbf{Obj 1:}} With the objective function of (\ref{Obj 1}) and the MMSE matrix in (\ref{MSE_Matrix_M}), we have
\begin{align}
\label{App_Inequ_1}
{\boldsymbol f}( {\bf{I}}_N-{\bf{Q}}_0^{\rm{H}}{\boldsymbol \Theta}{\bf{Q}}_0 )&={\rm{Tr}}({\bf{W}})-{\rm{Tr}}({\bf{W}}{\bf{Q}}_0^{\rm{H}}{\boldsymbol \Theta}{\bf{Q}}_0) \ge\underbrace{{\rm{Tr}}({\bf{W}})-\sum_{i=1}^N\lambda_i({\bf{W}})\lambda_i({\boldsymbol \Theta})}_{\triangleq {\boldsymbol g}({\boldsymbol \lambda}({\boldsymbol \Theta}))}
\end{align}where the inequality follows from the fact that for two positive semi-definite matrices ${\bf W}$ and ${\boldsymbol \Theta}$, ${\bf{Tr}}({\bf W }{\bf{Q}}_0^{\rm{H}}{\boldsymbol \Theta}{\bf{Q}}_0)\le \sum_i \lambda_i({\bf W})\lambda_i({\bf{Q}}_0^{\rm{H}}{\boldsymbol \Theta}{\bf{Q}}_0)$ with $\lambda_i({\boldsymbol Z})$ denoting the $i^{\rm{th}}$ largest eigenvalue of ${\boldsymbol Z}$. Furthermore, the second equality in (\ref{App_Inequ_1}) holds when ${\bf{Q}}_0^{\rm{H}}{\boldsymbol \Theta}{\bf{Q}}_0={\bf{U}}_{\bf{W}}{\rm{diag}}({\boldsymbol \lambda}({\boldsymbol \Theta})){\bf{U}}_{\bf{W}}^{\rm{H}}$ where $
{\boldsymbol \lambda}({\boldsymbol \Theta})=[{\lambda}_1({\boldsymbol \Theta}), \cdots, {\lambda}_N({\boldsymbol \Theta})]^{\rm{T}}$ and ${\bf{U}}_{\bf{W}}$ is the unitary matrix containing the eigenvectors of ${\bf{W}}$ as columns \cite{Marshall79}. It implies that the optimal value of ${\boldsymbol f}( {\bf{I}}_N-{\bf{Q}}_0^{\rm{H}}{\boldsymbol \Theta}{\bf{Q}}_0 )$ is ${\boldsymbol g}({\boldsymbol \lambda}({\boldsymbol \Theta}))$ and is achieved when ${\bf{Q}}_0^{\rm{H}}{\boldsymbol \Theta}{\bf{Q}}_0={\bf{U}}_{\bf{W}}{\rm{diag}}({\boldsymbol \lambda}({\boldsymbol \Theta})){\bf{U}}_{\bf{W}}^{\rm{H}}$.

Using the Lemma \textbf{2.A.2} \cite{Marshall79} and the definition of ${\boldsymbol g}({\boldsymbol \lambda}({\boldsymbol \Theta}))$ in (\ref{App_Inequ_1}), it can be easily found that ${\boldsymbol g}({\boldsymbol \lambda}({\boldsymbol \Theta}))$ is a Schur-concave function with respect to ${\boldsymbol \lambda}({\boldsymbol \Theta})$. Furthermore, for two vectors ${\bf{v}}\le{\bf{u}}$ (i.e., $v_i \le u_i$), from the definition of ${\boldsymbol g}({\boldsymbol \lambda}({\boldsymbol \Theta}))$ in (\ref{App_Inequ_1}), it can be concluded that ${\boldsymbol g}({\bf{v}})\ge {\boldsymbol g}({\bf{u}})$. It means that ${\boldsymbol g}(\bullet)$ is a decreasing function.

\noindent \underline{\textbf{Obj 2:}}  For the second objective function given by (\ref{Obj 2}), it is directly obtained that
\begin{align}
\label{App_Inequ_2}
{\boldsymbol f}( {\bf{I}}_N-{\bf{Q}}_0^{\rm{H}}{\boldsymbol \Theta}{\bf{Q}}_0 )&={\rm{log}}|{\bf{I}}_N-{\bf{Q}}_0^{\rm{H}}{\boldsymbol \Theta}{\bf{Q}}_0|=\underbrace{\sum_{i=1}^N {\rm{log}}[1-\lambda_i({\boldsymbol \Theta})]}_{\triangleq {\boldsymbol g}({\boldsymbol \lambda}({\boldsymbol \Theta}))}.
\end{align}
Obviously, the above equality holds unconditionally and thus the objective function ${\boldsymbol f}( {\bf{I}}_N-{\bf{Q}}_0^{\rm{H}}{\boldsymbol \Theta}{\bf{Q}}_0 )$ is independent of ${\bf{Q}}_0$. It follows from the optimization problem (\ref{Opt_X}) that ${\bf{Q}}_0$ can take any arbitrary unitary matrix since it is only involved in the constraint of ${\bf{Q}}_0^{\rm{H}}{\bf{Q}}_0={\bf{I}}$. Therefore, ${\bf{Q}}_0^{\rm{H}}{\boldsymbol \Theta}{\bf{Q}}_0={\bf{U}}_{\rm{Arb}}{\rm{diag}}({\boldsymbol \lambda}({\boldsymbol \Theta})){\bf{U}}_{\rm{Arb}}^{\rm{H}}$ with ${\bf{U}}_{\rm{Arb}}$ being an arbitrary unitary matrix always holds. Meanwhile, the optimal value of ${\boldsymbol f}( {\bf{I}}_N-{\bf{Q}}_0^{\rm{H}}{\boldsymbol \Theta}{\bf{Q}}_0 )$ can always be written as ${\boldsymbol g}({\boldsymbol \lambda}({\boldsymbol \Theta}))$. Based on the Lemma \textbf{2.A.2} \cite{Marshall79} and the definition of ${\boldsymbol g}({\boldsymbol \lambda}({\boldsymbol \Theta}))$ in (\ref{App_Inequ_2}), it can also be proved that ${\boldsymbol g}({\boldsymbol \lambda}({\boldsymbol \Theta}))$ is a decreasing Schur-concave function with respect to ${\boldsymbol \lambda}({\boldsymbol \Theta})$.

\noindent \underline{\textbf{Obj 3:}} For the diagonal elements of the positive semi-definite matrix ${\boldsymbol \Phi}_{\rm{MMSE}}(\{{\bf{P}}_k\}_{k=1}^K)={\bf{I}}_N-{\bf{Q}}_0^{\rm{H}}{\boldsymbol \Theta}{\bf{Q}}_0$, we have the following majorization relationship \cite{Marshall79}: \begin{align}
\label{APP_64}
{\rm{d}}({\bf{I}}_N-{\bf{Q}}_0^{\rm{H}}{\boldsymbol \Theta}{\bf{Q}}_0) \succ{\bf{1}}_N-({\sum}_{i=1}^N{\lambda}_i({\boldsymbol \Theta})/N )\otimes{\bf{1}}_N
\end{align}
where the equality holds if and only if $[{\bf{Q}}_0^{\rm{H}}{\boldsymbol \Theta}{\bf{Q}}_0]_{i,i}={\sum}_{i=1}^N{\lambda}_i({\boldsymbol \Theta})/N $, and ${\bf{1}}_N$ is the $N \times 1$ all-one vector.

For the third objective function in (\ref{Obj 3}), as ${\boldsymbol \psi}_1(\bullet)$ is increasing and Schur-convex, the objective function in (\ref{Opt_X}) satisfies \cite{Palomar03}
\begin{align}
\label{App_Inequ_3}
{\boldsymbol f}( {\bf{I}}_N-{\bf{Q}}_0^{\rm{H}}{\boldsymbol \Theta}{\bf{Q}}_0 )&= {\boldsymbol \psi}_1({\rm{d}}({\bf{I}}_N-{\bf{Q}}_0^{\rm{H}}{\boldsymbol \Theta}{\bf{Q}}_0))\ge\underbrace{{\boldsymbol \psi}_1\left({\bf{1}}_N-({\sum}_{i=1}^N{\lambda}_i({\boldsymbol \Theta})/N )\otimes{\bf{1}}_N\right)}_{\triangleq{\boldsymbol g}[{\boldsymbol{\lambda}}({\boldsymbol \Theta})]},
\end{align}
with equality if and only if $[{\bf{Q}}_0^{\rm{H}}{\boldsymbol \Theta}{\bf{Q}}_0]_{i,i}={\sum}_{i=1}^N{\lambda}_i({\boldsymbol \Theta})/N $. As shown in \cite{Palomar03}, when ${\bf{Q}}_0^{\rm{H}}{\boldsymbol \Theta}{\bf{Q}}_0={\bf{Q}}_{\rm{DFT}}{\rm{diag}}({\boldsymbol \lambda}({\boldsymbol \Theta})){\bf{Q}}_{\rm{DFT}}^{\rm{H}}$ where ${\bf{Q}}_{\rm{DFT}}$ is a DFT matrix, ${\bf{Q}}_0^{\rm{H}}{\boldsymbol \Theta}{\bf{Q}}_0$ has identical diagonal elements. It follows that when ${\bf{Q}}_0^{\rm{H}}{\boldsymbol \Theta}{\bf{Q}}_0={\bf{Q}}_{\rm{DFT}}{\rm{diag}}({\boldsymbol \lambda}({\boldsymbol \Theta})){\bf{Q}}_{\rm{DFT}}^{\rm{H}}$, the objective function ${\boldsymbol f}( {\bf{I}}_N-{\bf{Q}}_0^{\rm{H}}{\boldsymbol \Theta}{\bf{Q}}_0 )$ will take minimum/optimal value of ${\boldsymbol g}({\boldsymbol{\lambda}}({\boldsymbol \Theta}))$.

Based on the fact that ${\boldsymbol \psi}_1(\bullet)$ is an increasing and Schur-convex function, it can be directly concluded from (\ref{App_Inequ_3}) that ${\boldsymbol g}({\boldsymbol{\lambda}}({\boldsymbol \Theta}))$ is a decreasing function of ${\boldsymbol{\lambda}}({\boldsymbol \Theta})$. Furthermore, based on the Lemma \textbf{2.A.2} \cite{Marshall79}, ${\boldsymbol \psi}_1(\bullet)$ is also a Schur-concave function of ${\boldsymbol{\lambda}}({\boldsymbol \Theta})$.

\noindent \underline{\textbf{Obj 4:}} Notice that for the positive semi-definite matrix ${\boldsymbol \Phi}_{\rm{MMSE}}(\{{\bf{P}}_k\}_{k=1}^K)={\bf{I}}_N-{\bf{Q}}_0^{\rm{H}}{\boldsymbol \Theta}{\bf{Q}}_0$, ${\rm{d}}({\bf{I}}_N-{\bf{Q}}_0^{\rm{H}}{\boldsymbol \Theta}{\bf{Q}}_0) \prec {\boldsymbol{\lambda}}({\bf{I}}_N-{\bf{Q}}_0^{\rm{H}}{\boldsymbol \Theta}{\bf{Q}}_0)$ \cite{Palomar03}. With the Schur-concave function of ${\boldsymbol \psi}_2(\bullet)$ in (\ref{Obj 4}), we have
\begin{align}
\label{App_Inequ_4}
{\boldsymbol f}( {\bf{I}}_N-{\bf{Q}}_0^{\rm{H}}{\boldsymbol \Theta}{\bf{Q}}_0 )&={\boldsymbol \psi}_2({\rm{d}}({\bf{I}}_N-{\bf{Q}}_0^{\rm{H}}{\boldsymbol \Theta}{\bf{Q}}_0))\ge \underbrace{{\boldsymbol \psi}_2([{\bf{1}}_N-{\boldsymbol{\lambda}}({\boldsymbol \Theta})])}_{\triangleq {\boldsymbol g}[{\boldsymbol{\lambda}}({\boldsymbol \Theta})]},
\end{align}
where the equality holds when $[{\bf{Q}}_0^{\rm{H}}{\boldsymbol \Theta}{\bf{Q}}_0]_{i,i}={ \lambda}_i({\boldsymbol \Theta})$. It is easy to see that when $
{\bf{Q}}_0^{\rm{H}}{\boldsymbol \Theta}{\bf{Q}}_0={\bf{I}}_N{\rm{diag}}({\boldsymbol \lambda}({\boldsymbol \Theta})){\bf{I}}_N$, the preceding condition is satisfied and then the objective function ${\boldsymbol f}( {\bf{I}}_N-{\bf{Q}}_0^{\rm{H}}{\boldsymbol \Theta}{\bf{Q}}_0 )$ achieves its minimum/optimal value of ${\boldsymbol g}({\boldsymbol{\lambda}}({\boldsymbol \Theta}))$.

Since ${\boldsymbol \psi}_2(\bullet)$ is increasing and Schur-concave, it is clear that ${\boldsymbol g}({\boldsymbol{\lambda}}({\boldsymbol \Theta}))$ is decreasing with respect to ${\boldsymbol{\lambda}}({\boldsymbol \Theta})$. Moreover, using \cite[3.A.6.a]{Marshall79}, it can be proved that ${\boldsymbol \psi}_2({\bf{1}}_N-{\boldsymbol{\lambda}}({\boldsymbol \Theta}))$ is also Schur-concave with respect to ${\boldsymbol{\lambda}}({\boldsymbol \Theta})$.

\section{Proof of Property 2}
\label{Appedix:3}
First notice that for two matrices ${\boldsymbol A}$ and ${\boldsymbol B}$ with compatible dimensions,
$\lambda_i({\boldsymbol A}{\boldsymbol B})=\lambda_i({\boldsymbol B}{\boldsymbol A})$ \cite[9.A.1.a]{Marshall79}. Together with the fact that for two positive semi-definite matrices ${\boldsymbol A}$ and ${\boldsymbol B}$, $\prod_{i=1}^{k}\lambda_{i}({\boldsymbol A}{\boldsymbol B})\le\prod_{i=1}^{k}\lambda_{i}({\boldsymbol A})\lambda_{i}({\boldsymbol B})$ \cite[9.H.1.a]{Marshall79}, we have\footnote{Note that in general ${\boldsymbol A}_k$ is not a square matrix.}
\begin{align}
& \prod_{i=1}^{k}\lambda_{i}({\boldsymbol A}_{1}^{\rm{H}}{\bf{Q}}_1^{\rm{H}}\cdots{\boldsymbol A}_{K}^{\rm{H}}{\bf{Q}}_K^{\rm{H}}{\bf{Q}}_K {\boldsymbol A}_{K}\cdots{\bf{Q}}_1{\boldsymbol A}_{1})  \nonumber \\
 & \ \ \ \ \ \ \le \prod_{i=1}^{k}\lambda_i({\boldsymbol A}_2^{\rm{H}}{\bf{Q}}_2^{\rm{H}}\cdots {\boldsymbol A}_K^{\rm{H}}{\bf{Q}}_K^{\rm{H}}{\bf{Q}}_K^{\rm{H}} {\boldsymbol A}_K\cdots {\boldsymbol A}_{2}{\bf{Q}}_2)\underbrace{\lambda_{i}({\bf{Q}}_1{\boldsymbol A}_1{\boldsymbol A}_1^{\rm{H}}{\bf{Q}}_1^{\rm{H}})}_{=\lambda_{i}({\boldsymbol A}_1{\boldsymbol A}_1^{\rm{H}})}\  \ k=1,\cdots,N.
\end{align}
Repeating this process, we have the following inequality:
\begin{align}
\label{App_Inequality}
&\prod_{i=1}^{k}\lambda_{i}({\boldsymbol A}_{1}^{\rm{H}}{\bf{Q}}_1^{\rm{H}}\cdots{\boldsymbol A}_{K}^{\rm{H}}{\bf{Q}}_K^{\rm{H}}{\bf{Q}}_K {\boldsymbol A}_{K}\cdots{\bf{Q}}_1{\boldsymbol A}_{1}) \le \prod_{i=1}^{k}\underbrace{\lambda_{i}({\boldsymbol A}_K^{\rm{H}}{\boldsymbol A}_K)\lambda_i({\boldsymbol A}_{K-1}^{\rm{H}}{\boldsymbol A}_{K-1})\cdots \lambda_i({\boldsymbol A}_1^{\rm{H}}{\boldsymbol A}_1)}_{\triangleq {\gamma}_{i}(\{{\bf{F}}_k\}_{k=1}^K)}.
\end{align}
Based on (\ref{App_Inequality}) and \textbf{5.A.2.b} in \cite{Marshall79}, we directly have
\begin{align}
&{\boldsymbol \lambda}({\boldsymbol A}_{1}^{\rm{H}}{\bf{Q}}_1^{\rm{H}}\cdots{\boldsymbol A}_{K}^{\rm{H}}{\bf{Q}}_K^{\rm{H}}{\bf{Q}}_K {\boldsymbol A}_{K}\cdots{\bf{Q}}_1{\boldsymbol A}_{1}) \prec_w [{\gamma}_{1}(\{{\bf{F}}_k\}_{k=1}^K)  \ \cdots {\gamma}_{N}(\{{\bf{F}}_k\}_{k=1}^K)]^{\rm{T}}\triangleq {\boldsymbol \gamma}(\{{\bf{F}}_k\}_{k=1}^K)
\end{align}
where $ {\bf{a}} \prec_w {\bf{b}}$ denotes that ${\bf{a}}$ is weakly  majorized by ${\bf{b}}$ \cite{Marshall79} and the equality holds if and only if the neighboring ${\boldsymbol{A}}_k$'s satisfy
\begin{align}
\label{72}
{\bf{Q}}_k={\bf{V}}_{{\boldsymbol{A}}_{k+1}}{\bf{U}}_{{\boldsymbol{A}}_{k}}^{\rm{H}}, \ \ k=1,\cdots,K-1
\end{align}
where ${\bf{U}}_{{\boldsymbol{A}}_k}$ and ${\bf{V}}_{{\boldsymbol{A}}_{k}}$ are defined based on the following singular value decomposition: ${\boldsymbol{A}}_k={\bf{U}}_{{\boldsymbol{A}}_k}
{\boldsymbol{\Lambda}}_{{\boldsymbol{A}}_k}{\bf{V}}_{{\boldsymbol{A}}_k}^{\rm{H}} \ \ \text{with} \  \ {\boldsymbol{\Lambda}}_{{\boldsymbol{A}}_k} \searrow$. As ${\boldsymbol g}(\bullet)$ is a decreasing and Schur-concave function, we have \cite{Marshall79}
\begin{align}
& {\boldsymbol g}[{\boldsymbol \lambda}({\boldsymbol \Theta})]\ge {\boldsymbol g}[{\boldsymbol \gamma}(\{{\bf{F}}_k\}_{k=1}^K)]
\end{align}
with equality if and only if (\ref{72}) holds. Finally, based on the definition of ${\boldsymbol A}_k$ in (\ref{MSE_Matrix_M}), using the matrix inversion lemma,  the following equality holds:
\begin{align}
{\boldsymbol A}_k^{\rm{H}}{\boldsymbol A}_k={\bf{I}}_{M_k}-({\bf{F}}_k^{\rm{H}}{\bf{\bar H}}_k^{\rm{H}}{\bf{K}}_{{\bf{F}}_k}^{-1}{\bf{\bar H}}_k{\bf{F}}_k+{\bf{I}}_{M_k})^{-1}.
\end{align}
It follows that $\lambda_i({\boldsymbol A}_k^{\rm{H}}{\boldsymbol A}_k)={{ \lambda}_i({\bf{F}}_k^{\rm{H}}{\bf{\bar H}}_k^{\rm{H}}{\bf{K}}_{{\bf{F}}_k}^{-1}{\bf{\bar H}}_k{\bf{F}}_k)}/[{{1+{ \lambda}_i({\bf{F}}_k^{\rm{H}}{\bf{\bar H}}_k^{\rm{H}}{\bf{K}}_{{\bf{F}}_k}^{-1}{\bf{\bar H}}_k{\bf{F}}_k)}}]$.
Based on this result, ${\gamma}_{i}(\{{\bf{F}}_k\}_{k=1}^K)$ in (\ref{App_Inequality}) equals
\begin{align}
\label{73}
{\gamma}_{i}(\{{\bf{F}}_k\}_{k=1}^K)=\prod_{k=1}^K\frac{{ \lambda}_i({\bf{F}}_k^{\rm{H}}{\bf{\bar H}}_k^{\rm{H}}{\bf{K}}_{{\bf{F}}_k}^{-1}{\bf{\bar H}}_k{\bf{F}}_k)}{{1+{ \lambda}_i({\bf{F}}_k^{\rm{H}}{\bf{\bar H}}_k^{\rm{H}}{\bf{K}}_{{\bf{F}}_k}^{-1}{\bf{\bar H}}_k{\bf{F}}_k)}}.
\end{align}

\section{Proof of Property 3}
\label{Appedix:4}
Suppose at the moment that using the optimal ${\bf{F}}_{k}$, denoted by ${\bf{F}}_{k,{\rm{opt}}}$, transmission is not at the maximum power, i.e., ${\rm{Tr}}({\bf{F}}_{k,{\rm{opt}}}{\bf{F}}_{k,{\rm{opt}}}^{\rm{H}})<P_k$; then we have
$a\triangleq \sqrt{{P_k}/{{\rm{Tr}}({\bf{F}}_{k,{\rm{opt}}}{\bf{F}}_{k,{\rm{opt}}}^{\rm{H}})}}>1$. Defining ${\bf{\hat F}}_{k}\triangleq a{\bf{F}}_{k,\rm{opt}}$, it follows that
\begin{align}
\label{Power}
&{\bf{\hat F}}_{k}^{\rm{H}}{\bf{\bar H}}_k^{\rm{H}}{\bf{K}}_{{\bf{\hat F}}_{k}}^{-1}
{\bf{\bar H}}_k{\bf{\hat F}}_{k}\nonumber \\
=&{\bf{F}}_{k,{\rm{opt}}}^{\rm{H}}{\bf{\bar H}}_k^{\rm{H}}({\rm{Tr}}
({\bf{F}}_{k,{\rm{opt}}}
{\bf{F}}_{k,{\rm{opt}}}^{\rm{H}}
{\boldsymbol \Psi}_k){\boldsymbol \Sigma}_k+\sigma_{n_k}^2/a^2{\bf{I}})^{-1}{\bf{\bar H}}_k{\bf{F}}_{k,{\rm{opt}}} \nonumber \\
\succeq & {{\bf{F}}_{k,{\rm{opt}}}^{\rm{H}}{\bf{\bar H}}_k^{\rm{H}}\underbrace{({\rm{Tr}}({\bf{F}}_{k,{\rm{opt}}}
{\bf{F}}_{k,{\rm{opt}}}^{\rm{H}}
{\boldsymbol \Psi}_k){\boldsymbol \Sigma}_k+\sigma_{n_k}^2{\bf{I}})^{-1}}_{{\bf{K}}^{-1}_{{\bf{F}}_{k,{\rm{opt}}}}}{\bf{\bar H}}_k{\bf{F}}_{k,{\rm{opt}}}}.
\end{align} Note that ${\boldsymbol A} \succeq {\boldsymbol B}$ means that $\lambda_i({\boldsymbol A} ) \ge \lambda_i({\boldsymbol B})$ for all $i$, and therefore (\ref{Power}) implies
\begin{align}
\label{lambda}
{ \lambda}_i({\bf{\hat F}}_{k}^{\rm{H}}{\bf{\bar H}}_k^{\rm{H}}{\bf{K}}_{{\bf{\hat F}}_{k}}^{-1}
{\bf{\bar H}}_k{\bf{\hat F}}_{k})\ge { \lambda}_i({\bf{ F}}_{k,{\rm{opt}}}^{\rm{H}}{\bf{\bar H}}_k^{\rm{H}}{\bf{K}}_{{\bf{F}}_{k,{\rm{opt}}}}^{-1}
{\bf{\bar H}}_k{\bf{F}}_{k,{\rm{opt}}}).
\end{align} Moreover, it is clear from the definition of  ${\gamma}_{i}(\{{\bf{F}}_k\}_{k=1}^K)$ in (\ref{Property2_gamma}) that ${\gamma}_{i}(\{{\bf{F}}_k\}_{k=1}^K)$ is an increasing function of ${\lambda}_i({\bf{F}}_k^{\rm{H}}{\bf{\bar H}}_k^{\rm{H}}{\bf{K}}_{{\bf{F}}_k}^{-1}{\bf{\bar H}}_k{\bf{F}}_k)$. It then follows that ${\boldsymbol \gamma}(\{{\bf{F}}_k={\bf{\hat F}}_{k}\}_{k=1}^K))\ge {\boldsymbol \gamma}(\{{\bf{F}}_k={\bf{\hat F}}_{k,{\rm{opt}}}\}_{k=1}^K)$. Together with the fact that ${\boldsymbol g}(\bullet)$ is a decreasing function,  it is concluded that ${\boldsymbol g}[{\boldsymbol \gamma}(\{{\bf{F}}_k={\bf{\hat F}}_{k}\}_{k=1}^K))]\le {\boldsymbol g}[ {\boldsymbol \gamma}(\{{\bf{F}}_k={\bf{\hat F}}_{k,{\rm{opt}}}\}_{k=1}^K)]$. It is obvious that this result contradicts the optimality of ${\bf{F}}_{k,{\rm{opt}}}$, and therefore a necessary condition for the optimal ${\bf{F}}_k$ is ${\rm{Tr}}({\bf{F}}_{k}{\bf{F}}_{k}^{\rm{H}})=P_k$. Furthermore, when ${\rm{Tr}}({\bf{F}}_k{\bf{F}}_k^{\rm{H}})=P_k$, the following equality holds:
\begin{align}
\label{App_81}
{\rm{Tr}}[{\bf{F}}_k{\bf{F}}_k^{\rm{H}}(\alpha_k P_k{\boldsymbol
\Psi}_k+\sigma_{n_k}^2{\bf{I}})]&=\alpha_k P_k{\rm{Tr}}({\bf{F}}_k{\bf{F}}_k^{\rm{H}}{\boldsymbol \Psi}_k)+\sigma_{n_k}^2\underbrace
{{\rm{Tr}}({\bf{F}}_k{\bf{F}}_k^{\rm{H}})}_{=P_k}\nonumber \\&=\alpha_k P_k{\rm{Tr}}({\bf{F}}_k{\bf{F}}_k^{\rm{H}}{\boldsymbol \Psi}_k)+\sigma_{n_k}^2P_k.
\end{align}Defining $\eta_{f_k}=\alpha_k{\rm{Tr}}({\bf{F}}_k{\bf{F}}_k^{\rm{H}}{\boldsymbol \Psi}_k)+\sigma_{n_k}^2 $ with $\alpha_k={\rm{Tr}}({\boldsymbol \Sigma}_k)/M_k$, (\ref{App_81}) can be rewritten as ${\rm{Tr}}[{\bf{F}}_k{\bf{F}}_k^{\rm{H}}(\alpha_k P_k{\boldsymbol
\Psi}_k+\sigma_{n_k}^2{\bf{I}})]=P_k\eta_{f_k}$. In other words, the power constraint ${\rm{Tr}}({\bf{F}}_k{\bf{F}}_k^{\rm{H}})=P_k$ is equivalent to
\begin{align}
{\rm{Tr}}[{\bf{F}}_k{\bf{F}}_k^{\rm{H}}(\alpha_k P_k{\boldsymbol
\Psi}_k+\sigma_{n_k}^2{\bf{I}})]/\eta_{f_k}=P_k.
\end{align}

\section{Proof of Property 4}
\label{Appedix:5}

\noindent \underline{\textbf{Problem reformulation:}} As shown in (\ref{Property2_gamma}), ${\gamma}_i(\{{\bf{F}}_k\}_{k=1}^K)$ is a complicated function of ${\boldsymbol \lambda}({\bf{F}}_k^{\rm{H}}{\bf{\bar H}}_k^{\rm{H}}{\bf{K}}_{{\bf{F}}_k}^{-1}{\bf{\bar H}}_k{\bf{F}}_k)$. Clearly, ${\bf{F}}_k$ appears in multiple positions. In particular, ${\bf{K}}_{{\bf{F}}_k}$ is a function of ${\bf{F}}_k$ which complicates the derivation of optimal solutions. In order to simplify the problem,  ${\boldsymbol \lambda}({\bf{F}}_k^{\rm{H}}{\bf{\bar H}}_k^{\rm{H}}{\bf{K}}_{{\bf{F}}_k}^{-1}{\bf{\bar H}}_k{\bf{F}}_k)$ is reformulated as
\begin{align}
\label{APP_BB}
{\boldsymbol \lambda}({\bf{F}}_k^{\rm{H}}{\bf{\bar H}}_k^{\rm{H}}{\bf{K}}_{{\bf{F}}_k}^{-1}{\bf{\bar H}}_k{\bf{F}}_k)
=&{\boldsymbol \lambda}[{\bf{\tilde F}}_k^{\rm{H}}(\alpha_k P_k{\boldsymbol
\Psi}_k+\sigma_{n_k}^2{\bf{I}}_{N_k})^{-1/2}{\bf{\bar H}}_k^{\rm{H}}({\bf{K}}_{{\bf{F}}_k}/\eta_{f_k})^{-1/2}\nonumber \\
& \times \underbrace{ ({\bf{K}}_{{\bf{F}}_k}/\eta_{f_k})^{-1/2}{\bf{\bar H}}_k(\alpha_k P_k{\boldsymbol
\Psi}_k+\sigma_{n_k}^2{\bf{I}}_{N_k})^{-1/2}}_{\triangleq {\boldsymbol {\mathcal{H}}}_k}{\bf{\tilde F}}_k],
\end{align}where ${\bf{\tilde F}}_k$ is defined as
\begin{align}
\label{App_equ_X}
{\bf{\tilde F}}_k={1}/{\sqrt{\eta_{f_k}}}(\alpha_k P_k{\boldsymbol
\Psi}_k+\sigma_{n_k}^2{\bf{I}}_{N_k})^{1/2}{\bf{F}}_k.
\end{align}The right hand side of (\ref{APP_BB}) is  easier to handle than the left hand side. This is because when ${\boldsymbol{ \Psi}}_k\propto {\bf{I}}_{N_k}$ or ${\boldsymbol \Sigma}_k\propto{\bf{I}}_{M_k}$, ${\boldsymbol {\mathcal{H}}}_k$ is independent of ${\bf{\tilde F}}_k$. In the following, we will prove this in detail.

It is obvious that ${\boldsymbol {\mathcal{H}}}_k$ being independent of ${\bf{\tilde F}}_k$ is equivalent to ${\bf{K}}_{{\bf{F}}_k}/\eta_{f_k}$ being independent of ${\bf{\tilde F}}_k$. First consider ${\boldsymbol{ \Psi}}_k\propto{\bf{I}}_{N_k}$, i.e., ${\boldsymbol{ \Psi}}_k=\beta_k{\bf{I}}_{N_k}$. With the definitions of ${\bf{K}}_{{\bf{F}}_k}$ in (\ref{F_definition}) and $\eta_{f_k}$, ${\bf{K}}_{{\bf{F}}_k}/\eta_{f_k}$ equals
\begin{align}
{\bf{K}}_{{\bf{F}}_k}/\eta_{f_k}
&=[{\beta_k{\rm{Tr}}({\bf{F}}_k{\bf{F}}_k^{\rm{H}}){\boldsymbol \Sigma}_k+{\sigma}_{n_k}^2{\bf{I}}}_{M_k}]/[{\beta_k \alpha_k{\rm{Tr}}({\bf{F}}_k{\bf{F}}_k^{\rm{H}})+\sigma_{n_k}^2}]\nonumber \\
&=({\beta_k P_k}{\boldsymbol \Sigma}_k+{\sigma_{n_k}^2}{\bf{I}}_{M_k})/({\alpha_k \beta_k P_k+\sigma_{n_k}^2}),
\end{align}where the second equality is based on the fact that ${\rm{Tr}}({\bf{F}}_k{\bf{F}}_k^{\rm{H}})=P_k$ for the optimal ${\bf{F}}_k$. On the other hand, when ${\boldsymbol{ \Sigma}}_k \propto{\bf{I}}_{M_k}$ (i.e, ${\boldsymbol{ \Sigma}}_k=\alpha_k{\bf{I}}_{M_k}$), ${\bf{ K}}_{{\bf{F}}_k}/\eta_{f_k}$ equals
\begin{align}
{\bf{K}}_{{\bf{F}}_k}/\eta_{f_k}&= [\alpha_k{\rm{Tr}}({\bf{F}}_k{\bf{F}}_k^{\rm{H}}{\boldsymbol \Psi}_k){\bf{I}}_{N_k}+{\sigma}_{n_k}^2{\bf{I}}_{N_k}]/ [{ \alpha_k[{\rm{Tr}}({\bf{F}}_k{\bf{F}}_k^{\rm{H}}{\boldsymbol \Psi}_k)+\sigma_{n_k}^2}]\nonumber \\
& ={\bf{I}}_{N_k}.
\end{align} Therefore, when ${\boldsymbol{ \Psi}}_k\propto {\bf{I}}_{N_k}$ or ${\boldsymbol \Sigma}_k\propto{\bf{I}}_{M_k}$, ${\bf{K}}_{{\bf{F}}_k}/\eta_{f_k}$ is independent of ${\bf{\tilde F}}_k$.

Using the substitution (\ref{App_equ_X}), the optimization problem (\ref{Opt_X_Final}) is reformulated as
 \begin{align}
\label{OPT_X_Apped}
& \min_{{\bf{\tilde F}}_k} \ \ {\boldsymbol g} [{\boldsymbol \gamma}(\{{\bf{\tilde F}}_k\}_{k=1}^K)] \nonumber \\
& \  {\rm{s.t.}} \ \ \ {\gamma}_{i}(\{{\bf{\tilde F}}_k\}_{k=1}^K)=\prod_{k=1}^K \frac{{ \lambda}_i({\bf{\tilde F}}_k^{\rm{H}}{\boldsymbol{\mathcal{H}}}_k^{\rm{H}}{\boldsymbol{\mathcal{H}}}_k
 {\bf{\tilde F}}_k)}{1+{ \lambda}_i({\bf{\tilde F}}_k^{\rm{H}}{\boldsymbol{\mathcal{H}}}_k^{\rm{H}}
 {\boldsymbol{\mathcal{H}}}_k{\bf{\tilde F}}_k)} \nonumber \\
 & \ \ \ \ \ \ \ \ \  {\rm{Tr}}({\bf{\tilde F}}_k{\bf{\tilde F}}_k^{\rm{H}})=P_k.
\end{align}

\noindent \underline{\textbf{Structure of optimal ${\bf{\tilde F}}_k$:}} For the optimal ${\bf{\tilde F}}_k$, denoted as ${\bf{\tilde F}}_{k,{\rm{opt}}}$, based on the following singular value decompositions:
\begin{align}
{\boldsymbol{\mathcal{H}}}_k{\bf{\tilde F}}_{k,{\rm{opt}}}&={\bf{U}}_{{\boldsymbol M}_{k}}
{\boldsymbol \Lambda}_{{\boldsymbol M}_{k}}{\bf{V}}_{{\boldsymbol M}_{k}}^{\rm{H}} \ \ \text{with} \ \ {\boldsymbol \Lambda}_{{\boldsymbol M}_{k}} \searrow \ \ {\text{and}} \ \
{\boldsymbol{\mathcal{H}}}_k&
={\bf{U}}_{{\boldsymbol{\mathcal{H}}}_k}{\boldsymbol \Lambda}_{{\boldsymbol{\mathcal{H}}}_k}{\bf{V}}_{{\boldsymbol{\mathcal{H}}}_k}^{\rm{H}} \ \ \text{with} \ \  {\boldsymbol \Lambda}_{{\boldsymbol{\mathcal{H}}}_k} \searrow,
\end{align}
we can construct a matrix ${\bf{\bar F}}_k$,
\begin{align}
\label{Apped_X}
&{\bf{\bar F}}_k={\bf{V}}_{{\boldsymbol{\mathcal{H}}}_k}
{\boldsymbol \Lambda}_{{\bf{X}}_k}{\bf{V}}_{{\boldsymbol M}_{k}}^{\rm{H}},
\end{align}
where ${\boldsymbol \Lambda}_{{\bf{X}}_k}$ is an unknown diagonal matrix with the same rank as ${\boldsymbol \Lambda}_{{\boldsymbol M}_{k}}$ and $
{\boldsymbol \Lambda}_{{\boldsymbol{\mathcal{H}}}_k}{\boldsymbol \Lambda}_{{\bf{X}}_k}/b
={\boldsymbol \Lambda}_{{\boldsymbol M}_{k}}$, and the scalar $b$ is chosen to make ${\rm{Tr}}({\bf{\bar F}}_k {\bf{\bar F}}_k^{\rm{H}})=P_k$ hold.

Because ${\boldsymbol {\mathcal{H}}}_k$ is independent of the unknown variable ${\bf{\tilde F}}_k$, using Lemma 12 in \cite{Palomar03}, we have
\begin{align}
{\bf{\bar F}}_k^{\rm{H}}{\boldsymbol{\mathcal{H}}}_k^{\rm{H}}{\boldsymbol{\mathcal{H}}}_k{\bf{\bar F}}_k
\succeq {\bf{\tilde F}}_{k,{\rm{opt}}}^{\rm{H}}{\boldsymbol{\mathcal{H}}}_k^{\rm{H}}{\boldsymbol{\mathcal{H}}}_k
{\bf{\tilde F}}_{k,{\rm{opt}}}.
\end{align}
Taking eigenvalues of both sides, we have $\lambda_i({\bf{\bar F}}_k^{\rm{H}}{\boldsymbol{\mathcal{H}}}_k^{\rm{H}}{\boldsymbol{\mathcal{H}}}_k{\bf{\bar F}}_k)\ge \lambda_i({\bf{\tilde F}}_{k,{\rm{opt}}}^{\rm{H}}{\boldsymbol{\mathcal{H}}}_k^{\rm{H}}{\boldsymbol{\mathcal{H}}}_k
{\bf{\tilde F}}_{k,{\rm{opt}}})$ \cite{Marshall79}. Since ${\gamma}_{i}(\{{\bf{\tilde F}}_k\}_{k=1}^K)$ is an increasing function of $\lambda_i({\bf{\tilde F}}_{k}^{\rm{H}}{\boldsymbol{\mathcal{H}}}_k^{\rm{H}}{\boldsymbol{\mathcal{H}}}_k
{\bf{\tilde F}}_{k})$, we directly have that ${\boldsymbol \gamma}(\{{{\bf{\tilde F}}_k={\bf{{\bar F}}}_{k}}\}_{k=1}^K)
\ge {\boldsymbol \gamma}(\{{{\bf{\tilde F}}_k={\bf{{\tilde  F}}}_{k,{\rm{opt}}}}\}_{k=1}^K)$. Furthermore, as the objective function ${\boldsymbol g}(\bullet)$  of (\ref{OPT_X_Apped}) is a decreasing function,  we finally have ${\boldsymbol g} [{\boldsymbol \gamma}(\{{{\bf{\tilde F}}_k={\bf{{\bar F}}}_{k}}\}_{k=1}^K)]
\le {\boldsymbol g}[{\boldsymbol \gamma}(\{{{\bf{\tilde F}}_k={\bf{{\tilde  F}}}_{k,{\rm{opt}}}}\}_{k=1}^K)]$. Because ${{\bf{\tilde F}}_{k,{\rm{opt}}}}$ is the optimal solution, ${{\bf{\tilde F}}_{k,{\rm{opt}}}}$ must be in the form of ${\bf{\bar F}}_k$. Therefore, the structure of optimal ${\bf{\tilde F}}_k$ is given by (\ref{Apped_X}), i.e., ${\bf{\tilde F}}_{k,{\rm{opt}}}={\bf{V}}_{{\boldsymbol{\mathcal{H}}}_k}{\boldsymbol \Lambda}_{{\bf{X}}_k}{\bf{V}}_{{\boldsymbol M}_{k}}^{\rm{H}}$.

As the minimum dimension of ${\boldsymbol A}_k$ is $N$, on substituting (\ref{Apped_X}) into $\gamma_i(\{{\bf{F}}_k\}_{k=1}^K)$ (\ref{App_Inequality}), it can be seen that for the optimal solution only the $N \times N$ principal submatrix of ${\boldsymbol \Lambda}_{{\bf{X}}_k}$ can be nonzero, which is denoted as ${\boldsymbol { \Lambda}}_{{\boldsymbol{\mathcal{F}}}_k}$. As a result, ${\bf{\tilde F}}_{k,{\rm{opt}}}$ has the following structure:
\begin{align}
\label{Apped_X_OPT}
{\bf{\tilde F}}_{k,{\rm{opt}}}={\bf{V}}_{{\boldsymbol{\mathcal{H}}}_k,N}{\boldsymbol \Lambda}_{{\boldsymbol{\mathcal{F}}}_k}{\bf{V}}_{{\boldsymbol {M}}_{k},N}^{\rm{H}}.
\end{align}
It is clear that the values of ${\bf{V}}_{{\boldsymbol {M}}_{k}}$'s do not affect the values of $\lambda_i({\bf{\tilde F}}_{k}^{\rm{H}}{\boldsymbol{\mathcal{H}}}_k^{\rm{H}}{\boldsymbol{\mathcal{H}}}_k
{\bf{\tilde F}}_{k})$, the constraint ${\rm{Tr}}({\bf{\tilde F}}_k{\bf{\tilde F}}_k^{\rm{H}})=P_k$ and the objective function in the optimization problem (\ref{OPT_X_Apped}). Therefore, ${\bf{V}}_{{\boldsymbol {M}}_{k}}$ can be an arbitrary unitary matrix.

\noindent \underline{\textbf{Structure of optimal ${\bf{F}}_k$:}} Based on the relationship between ${\bf{F}}_k$ and ${\bf{\tilde F}}_k$ given in (\ref{App_equ_X}),
\begin{align}
\label{F_structure_opt}
 {\bf{F}}_{k,\rm{opt}}=\sqrt{\eta_{f_k}}(\alpha_k P_k{\boldsymbol
\Psi}_k+\sigma_{n_k}^2{\bf{I}}_{N_k})^{-1/2}{\bf{\tilde F}}_{k,\rm{opt}}.
\end{align} Putting the structure of ${\bf{F}}_{k,\rm{opt}}$ in (\ref{F_structure_opt}) into $\eta_{f_k}$ in (\ref{eta_f}), $\eta_{f_k}$ can be solved to be
\begin{align}
\label{App_eta}
\eta_{f_k}&={\sigma_{n_k}^2}/\{1-\alpha_k{\rm{Tr}}
[{\bf{V}}_{{\boldsymbol{\mathcal{H}}}_k,N}^{\rm{H}}(
\alpha_k P_k{\boldsymbol
\Psi}_k+\sigma_{n_k}^2{\bf{I}}_{N_k})^{-1/2}{\boldsymbol
\Psi}_k(\alpha_k P_k{\boldsymbol
\Psi}_k+\sigma_{n_k}^2{\bf{I}}_{N_k})^{-1/2}
{\bf{V}}_{{\boldsymbol{\mathcal{H}}}_k,N}{\boldsymbol { \Lambda}}_{{\boldsymbol{\mathcal{F}}}_k}^2]\}.
\end{align} Clearly in (\ref{App_eta}), $\eta_{f_k}$ is a function of ${\boldsymbol { \Lambda}}_{{\boldsymbol{\mathcal{F}}}_k}$ and it can be denoted as $\eta_{f_k}={ \xi}_k({\boldsymbol { \Lambda}}_{{\boldsymbol{\mathcal{F}}}_k})$ for clarification.

\end{document}